\documentclass[prc,aps,groupedaddress,superscriptaddress,nofootinbib,showpacs
,preprintnumbers,onecolumn]{revtex4}
\usepackage{graphicx}
\usepackage{amsfonts}
\usepackage{amssymb}
\usepackage{natbib}
\usepackage{dcolumn}
\usepackage{bm}
\usepackage{subeqnarray}

\def\lim{\mathop{\rm {lim}}}

\newcommand{\bwt}{\begin{widetext}}
\newcommand{\ewt}{\end{widetext}}
\newcommand{\beq}{\begin{equation}}
\newcommand{\eeq}{\end{equation}}
\newcommand{\bea}{\begin{eqnarray}}
\newcommand{\eea}{\end{eqnarray}}
\newcommand{\beas}{\begin{subeqnarray}}
\newcommand{\eeas}{\end{subeqnarray}}
\begin{document}


\title{Self-Consistent Quasi-Particle RPA for the Description of Superfluid
Fermi Systems.}

\author{A. Rabhi}
\email{rabhi@ipnl.in2p3.fr}
\affiliation{Laboratoire de Physique de la Mati\`ere Condens\'ee, 
Facult\'e des Sciences de Tunis, Campus Universitaire, Le Belv\'ed\`ere-1060, Tunisia} 
\affiliation{IPN-Lyon, 43Bd du 11 novembre 1918, F-69622 Villeurbanne Cedex, France}
\author{R. Bennaceur}
\email{bennaceur.raouf@inrst.rnrt.tn}
\affiliation{Laboratoire de Physique de la Mati\`ere Condens\'ee, 
Facult\'e des Sciences de Tunis, Campus Universitaire, Le Belv\'ed\`ere-1060, Tunisia} 
\author{G. Chanfray}
\email{chanfray@ipnl.in2p3.fr}
\affiliation{IPN-Lyon, 43Bd du 11 novembre 1918, F-69622 Villeurbanne Cedex, France}
\author{P. Schuck}
\email{schuck@ipno.in2p3.fr}
\affiliation{Institut de Physique Nucl\'eaire, IN2P3-CNRS, Universit\'e Paris-Sud, 
F-91406 Orsay Cedex, France}

\date{\today}

\begin{abstract}
Self-Consistent Quasi-Particle RPA (SCQRPA) is for the first time applied to a more level 
pairing case. Various filling situations and values for the coupling  constant are considered. 
Very encouraging results in comparison with the exact solution of the model are obtained. The 
nature of the low lying mode in SCQRPA is identified. The strong reduction of the number 
fluctuation in SCQRPA vs BCS is pointed out. The transition from superfluidity to the normal 
fluid case is carefully investigated.
\end{abstract}

\maketitle

\section{Introduction}
\label{intro}  
One of the most spectacular quantum phenomena is the transition to the 
supraconducting or superfluid state in interacting Fermi systems. This
happens e.g. in metals, liquid $^{3}He$, neutron stars, in finite
nuclei, and it is actively searched for in systems of magnetically trapped
atomic Fermions. In most of these systems the canonical mean field approach of
Bardeen, Cooper, and Schrieffer (BCS) with a couple of adjustable parameters
works astonishingly well.
However, in recent years there have been increasing attempts to describe the
pairing phenomenon on completely microscopic grounds. To our knowledge these
attempts have mostly been carried out for nuclear systems. This stems on the one hand from
the fact that phenomenological $NN$ forces are on the market which very well describe
the nucleon-nucleon phase shifts in all channels and in a wide range of
energies. On the other hand the physics of neutron stars makes quantitative
predictions of the pairing phenomenon in neutron matter indispensable, since
superfluidity of neutrons stars manifests itself only quite indirectly through e.g.
the phenomenon of neutron star glitches.
The microscopic approaches to pairing, starting from a bare two body
interaction, are not very numerous. The simplest one is based on BCS theory, using
however, in the gap equation the bare force and for the single
particle dispersion the one given by Br\"uckner theory. In this way one obtains
e.g. gap values in the $^{1}S_{0}$ channel for neutron-neutron pairing which 
in infinite matter, as a function of the Fermi momentum $k_{F}$, have a typical bell
shaped form roughly dropping to zero around $k_{F}=1.3 fm^{-1}$ and culminating 
at $k_{F}=0.8 fm^{-1}$ to values of $\Delta =2.5 - 3.0 MeV$ for neutron and nuclear
matter respectively. This rather elementary approach has been extended in the
past in various ways. The most ambitious procedure is probably the so-called
correlated basis function approach~\cite{b1}. However, more recently self-consistent
T-matrix approaches and extended Br\"uckner theories with rearrangement terms have
achieved a remarkable degree of sophistication~\cite{b2}. The screening of the interaction was
treated to lowest order in the density, resuming the RPA bubbles, in
introducing self-consistent Landau parameters~\cite{b3}. The outcome of all these
investigations inevitably leads to a quite substantial reduction of pairing in 
neutron matter but also in symmetric nuclear matter. The global reduction
generally attains important values and often reaches factors close to three. 
Such small values of the gap in infinite matter, however, pose a problem. 
Employing the Local Density Approximation (LDA) to estimate from the infinite 
matter results the gap in finite nuclei~\cite{b4}, one reaches with the simplified
approach described above using the bare NN force quite reasonable gap values for
finite nuclei. Interestingly in the gap equation quite similar results are obtained
with the Gogny D${1}$S force~\cite{b5} using the same procedure. However, with such strongly
reduced gaps from the more sophisticated approaches mentioned above, one obtains much 
too small gaps in finite nuclei. 
Of course, this reasoning may be completely erroneous and the situation in
finite nuclei may be very different from infinite matter. Nevertheless we find
the above argumentation intriguing. On the other hand we know that pairing is an
extraordinarily subtle process and employing theories which are in one or the
other way uncontrolled may turn out to be a hazardous enterprise. In such a
situation it is probably wise to investigate the problem from different sides using
a variety of approaches.

In the past we have made very positive experience with an extension of RPA
theory which we called Self-Consistent RPA (SCRPA)~\cite{b6, b8, b7}. For instance, in a recent
work this theory has been applied to the exactly solvable many level pairing
model in the pre-critical regime and very good agreement with the exact results for
ground-state energy and the low lying part of the spectrum was found~\cite{b8, b7}. This
success has encouraged us to develop the SCRPA formalism also for the fully
developed superfluid regime. This is a not completely trivial extension of the
SCRPA and we here apply it for the first time to the two level pairing model. As
we will see the theory also gives very promising results in the superfluid phase.
Since the Self-Consistent Quasi-Particle RPA (SCQRPA), as
in general the SCRPA theory, can be derived from a variational principle, which
turns out to be very close to a Raleigh-Ritz variational theory, we believe 
that SCQRPA is a non perturbative approach going in a certain systematic way
beyond the mean field BCS theory, including in a self-consistent way
correlations and quantum fluctuations. It is our believe that this microscopic
approach can ultimately be used to calculate pairing properties of realistic Fermi
systems starting from the bare force.

It should be mentioned that extensions of RPA theory, based on the Equation of
Motion (EOM) method, have by now a quite long history. They, to a great deal,
have been developed in nuclear physics. It started out with the work of Hara who
included the ground-state correlation in the Fermion occupation numbers~\cite{b9}. More
systematic was the consequent work by Rowe and co-workers (see the review by D.
J. Rowe~\cite{c9}). The same theory was developed using the Green's function method by one
of the present authors~\cite{b6}. Independently the method was also proposed by Zimmermann
and G. R\"opke plus coworkers using a graphical construction~\cite{b10}. These authors
named their method Cluster-Hartree-Fock (CHF) and it is equivalent to
Self-Consistent RPA (SCRPA). The latter approach has recently been further
developed by Dukelsky and Schuck in a series of papers~\cite{b8,b7,b15,b11,b12,b13}. 
However, also other authors contributed actively to the subject~\cite{b14}. A number of 
remarkable results have been obtained with SCRPA in non trivial models where comparison with 
exact solutions was possible~\cite{b7,b15}. For instance for the exactly solvable many 
level pairing model of Richardson~\cite{b16} SCRPA provides very accurate results for 
the ground state and the low lying part of the spectrum~\cite{b7,b8}.

In detail our paper is organized as follows : in section~\ref{model} the two level pairing 
model is introduced, in section~\ref{scqrpa} the SCQRPA formalism is presented, in 
section~\ref{results} numerical results are given and detailed discussions are presented. 
Comparison with other recent works is made in section~\ref{compar}, in section~\ref{senio} 
the question of the second constraint on the particle number variance is invoked and applied
to the Seniority model. In section~\ref{concl}, we will summarize the results and draw some 
conclusions. Finally, some useful mathematical relations and a second method for the calculation
of occupation numbers are given in the Appendices.

\section{The Model}
\label{model}  
The two-level pairing model is an exactly solvable model extensively employed in nuclear physics 
to test many-body approximations. It was first used to test the pp-RPA~\cite{c24} and its ability
to describe  ground-state correlations and vibrations in the normal phase as well as in the 
superfluid phase. The model is composed of two levels with equal degeneracy $2\Omega = 2J+1$ 
($J$ is the spin of each level) and an single-particle energy splitting $\epsilon$. The pairing 
Hamiltonian in this model space is
\beq
H=\frac{\epsilon}{2}\sum_{j}j \hat{N}_{j} - g\Omega \sum_{j
{j}^{\prime}} A_{j}^{\dagger} A_{{j}^{\prime}}, \; j =\pm 1
\label{a1}
\eeq
where ${j}$ takes the values $1$ for the upper level and $-1$ for the lower level. $\hat{N}_{j}$ 
and $A^{\dagger}_{j}$ are the number and monopole pair operators of the level $j$, respectively,
\beq
A^{\dagger}_{j}=\frac{1}{\sqrt{\Omega}}\sum_{m = 1}^{\Omega} 
a^{\dagger}_{j m}a^{\dagger}_{j \bar{m}}
\label{a2}
\eeq
and
\beq
\hat{N}_{j}=\sum_{m = 1}^{\Omega}(a^{\dagger}_{j m}a_{j m}+a^{\dagger}_{j \bar{m}}a_{j \bar{m}}).
\label{a3}
\eeq
where $a^{\dagger}_{j m}$ creates a particle in the level $j$ with spin projection $m$
and $a_{j \bar{m}}=(-1)^{J-m}a_{j -m}$. The operators obey the following commutations relations,
\bea
[A_{j}, A^{\dagger}_{j^{\prime}}]&=&\delta_{jj^{\prime}}\bigg(1-\frac{\hat{N}_{j}}{\Omega}\bigg), \cr
[\hat{N}_{j}, A^{\dagger}_{j^{\prime}}]&=&\delta_{jj^{\prime}}2A^{\dagger}_{j}, \cr
[\hat{N}_{j}, A_{j^{\prime}}]&=&-\delta_{jj^{\prime}}2A_{j},
\label{aa39}
\eea
thus, they define an $SU(2)$ algebra for each level and the two level model satisfies an 
$SU(2)\times SU(2)$ algebra.
 
For a system not at half filling, the normalized states in the Hilbert subspace of the monopole pairs are 
\bwt \beq
|n\rangle=\frac{{\Omega}^{\frac{\tilde{\Omega}}{2}}}{\Omega!}
\sqrt{\frac{(\Omega-\tilde{\Omega}+n)!(\Omega-n)!}{n!(\tilde{\Omega}-n)!}}
(A^{\dagger}_{1})^{n}(A^{\dagger}_{-1})^{\tilde{\Omega}-n}|0\rangle, 
\quad 0 \leqslant n \leqslant \tilde{\Omega}
\label{a6}
\eeq  \ewt
where $\tilde{\Omega}=\Omega$ leads to the half filling case, i.e. the lower level is filled for $g=0$.  
The matrix Hamiltonian is tridiagonal of dimension $\tilde{\Omega} + 1$, with matrix elements
\bwt \bea
h_{n,n}=\langle n|H|n\rangle=\epsilon (2 n -\tilde{\Omega}) -G (2 n \tilde{\Omega} - 2 n^{2}
+\tilde{\Omega} \Omega -\tilde{\Omega}^{2} +\tilde{\Omega}),\\
h_{n-1,n}=\langle n-1|H|n\rangle=-G \sqrt{n(\Omega-(n-1))(\Omega-\tilde{\Omega}+n)(\tilde{\Omega}-n+1)}
\label{a7}
\eea \ewt
where, $n$ is the number of pairs in the upper level and the number of particle is given by $N=2 
\tilde{\Omega}$.
\section{Self-consistent QRPA}
\label{scqrpa}
In a recent work~\cite{b7}  the SCRPA has been applied with very good success to the picket fence model 
in the non superfluid phase. The extension to the superfluid phase is slightly delicate and we here 
limit ourselves to the two level model, however considering arbitrary degeneracies and fillings of the 
levels. The objective in this section is to establish the equations for the
Self-Consistent Quasi-Particle RPA (SCQRPA). A first application of SCQRPA has
been performed in~\cite{b12} for the case of the seniority model (one-level pairing
model). We will again later come back to this model. Here we want to consider
the two level pairing model with arbitrary filling and coupling strength in the
SCQRPA approach which already more or less shows the full complexity of more 
realistic many level problems. As a first step we have to transform the constrained Hamiltonian  
\beq
H^{\prime}=H-\mu\hat{N},
\label{a8}
\eeq
where $\hat{N}$ is the full particle number operator, to quasi-particle operators
\beq
\pmatrix{\alpha^{\dagger}_{j m} \cr \alpha_{j \bar{m}}}=
\pmatrix{u_{j}&-v_{j}\cr
         v_{j}&u_{j}}
\pmatrix{a_{j m}^{\dagger} \cr a_{j \bar{m}}}
\label{a9}
\eeq 
\beq
\pmatrix{a_{j m}^{\dagger} \cr a_{j \bar{m}} }=
\pmatrix{u_{j}&v_{j}\cr
         -v_{j}&u_{j}}
\pmatrix{\alpha^{\dagger}_{j m} \cr \alpha_{j \bar{m}}}
\label{a10}
\eeq
with
\beq
u_{j}^{2}+v_{j}^{2}=1, \; j=\pm 1.
\label{a11}
\eeq
We define new quasi spin operators as
\beq
P^{\dagger}_{j}=\frac{1}{\sqrt{\Omega}}\sum_{m>0}\alpha^{\dagger}_{j m}
\alpha^{\dagger}_{j \bar{m}}, \quad P_{j}=(P_{j}^{\dagger})^{\dagger}
\label{a12}
\eeq
and the quasi-particle number operator in the level $j$ is given by,
\beq
\hat{N}_{q,j}=\sum_{m>0}(\alpha^{\dagger}_{j m}\alpha_{j m}
+\alpha^{\dagger}_{j \bar{m}}\alpha_{j \bar{m}}).
\label{a13}
\eeq
The quasi-particle operators obey the following commutations relations,
\bea
[P_{j}, P^{\dagger}_{j^{\prime}}]&=&\delta_{jj^{\prime}}
\bigg(1-\frac{\hat{N}_{q,j}}{\Omega}\bigg), \cr
[\hat{N}_{q,j}, P^{\dagger}_{j^{\prime}}]&=&\delta_{jj^{\prime}}2P^{\dagger}_{j}, \cr
[\hat{N}_{q,j}, P_{j^{\prime}}]&=&-\delta_{jj^{\prime}}2P_{j}.
\label{a39}
\eea
Then the Hamiltonian in the quasi-particle basis can be written as
\beq
H^{\prime}=H^{\prime}_{00}+H^{\prime}_{11}+H^{\prime}_{20}
+H^{\prime}_{22}+H^{\prime}_{31}+H^{\prime}_{40}+H^{\prime}_{11-11}
\label{a14}
\eeq
where
\bwt\bea
H^{\prime}_{00}&=&h_{0},\\
H^{\prime}_{11}&=&h_{1}\hat{N}_{q,1}+h_{-1}\hat{N}_{q,-1},\\
H^{\prime}_{20}&=&h_{2}(P^{\dagger}_{1}+P_{1})+h_{-2}(P^{\dagger}_{-1}+P_{-1}),\\
H^{\prime}_{22}&=&h_{3}P^{\dagger}_{1}P_{1}+h_{-3}P^{\dagger}_{-1}P_{-1}
                  +h_{4}(P^{\dagger}_{1}P_{-1}+P^{\dagger}_{-1}P_{1}),\\
H^{\prime}_{31}&=&h_{5}(P^{\dagger}_{1}\hat{N}_{q,1}+\hat{N}_{q,1}P_{1}) 
+h_{-5}(P^{\dagger}_{-1}\hat{N}_{q,-1}+\hat{N}_{q,-1}P_{-1}) \\
        &+&h_{6}(P^{\dagger}_{1}\hat{N}_{q,-1}+\hat{N}_{q,-1}P_{1}) 
        +h_{-6}(P^{\dagger}_{-1}\hat{N}_{q,1}+\hat{N}_{q,1}P_{-1}), \\
H^{\prime}_{40}&=&h_{7}(P^{\dagger}_{1}P^{\dagger}_{1}+P_{1}P_{1})
          +h_{-7}(P^{\dagger}_{-1}P^{\dagger}_{-1}+P_{-1}P_{-1})
          +h_{8}(P^{\dagger}_{1}P^{\dagger}_{-1}+P_{-1}P_{1}),\\
H^{\prime}_{11-11}&=&h_{9}\hat{N}_{q,1}^{2}+h_{-9}\hat{N}_{q,-1}^{2}
+h_{10}\hat{N}_{q,1}\hat{N}_{q,-1}
\label{a15}
\eea\ewt
and,
\bwt\bea
h_{0}&=&(\epsilon-2\mu) \Omega v_{1}^{2}-g\Omega (\Omega u_{1}^{2}v_{1}^{2}+v_{1}^{4})
-(\epsilon+2\mu) \Omega v_{-1}^{2}-g\Omega (\Omega u_{-1}^{2}v_{-1}^{2}+v_{-1}^{4}) \cr
&-&2g{\Omega}^{2}u_{1}v_{1}u_{-1}v_{-1},\cr
h_{1}&=&(\frac{\epsilon}{2}-\mu)(u_{1}^{2}-v_{1}^{2})+g\Omega (2
u_{1}^{2}v_{1}^{2}+\frac{v_{1}^{4}}{\Omega})+2g{\Omega}^{2}u_{1}v_{1}u_{-1}v_{-1}, \cr 
h_{-1}&=&-(\frac{\epsilon}{2}+\mu)(u_{-1}^{2}-v_{-1}^{2})+g\Omega (2
u_{-1}^{2}v_{-1}^{2}+\frac{v_{-1}^{4}}{\Omega})+2g{\Omega}^{2}u_{-1}v_{-1}u_{1}v_{1}, \cr
h_{2}&=&\sqrt{\Omega}u_{1}v_{1}(\epsilon-2\mu)
-g\Omega\bigg\{u_{1}v_{1}(u_{1}^{2}-v_{1}^{2})\sqrt{\Omega}
+\frac{2u_{1}v_{1}^{3}}{\sqrt{\Omega}}\bigg\}
-g\Omega\sqrt{\Omega}u_{-1}v_{-1}(u_{1}^{2}-v_{1}^{2}),
\cr
h_{-2}&=&-\sqrt{\Omega}u_{-1}v_{-1}(\epsilon+2\mu)
-g\Omega\bigg\{u_{-1}v_{-1}(u_{-1}^{2}-v_{-1}^{2})\sqrt{\Omega}
+\frac{2u_{-1}v_{-1}^{3}}{\sqrt{\Omega}}\bigg\} \cr
&-&g\Omega\sqrt{\Omega}u_{1}v_{1}(u_{-1}^{2}-v_{-1}^{2}), \cr
h_{3}&=&-g\Omega(u_{1}^{4}+v_{1}^{4}), \cr
h_{-3}&=&-g\Omega(u_{-1}^{4}+v_{-1}^{4}), \cr
h_{4}&=&-g\Omega(u_{1}^{2}u_{-1}^{2}+v_{1}^{2}v_{-1}^{2}), \cr
h_{5}&=&g\sqrt{\Omega}u_{1}v_{1}(u_{1}^{2}-v_{1}^{2}), \cr
h_{-5}&=&g\sqrt{\Omega}u_{-1}v_{-1}(u_{-1}^{2}-v_{-1}^{2}), \cr
h_{6}&=&g\sqrt{\Omega}u_{-1}v_{-1}(u_{1}^{2}-v_{1}^{2}), \cr
h_{-6}&=&g\sqrt{\Omega}u_{1}v_{1}(u_{-1}^{2}-v_{-1}^{2}), \cr
h_{7}&=&g\Omega u_{1}^{2}v_{1}^{2}, \cr
h_{-7}&=&g\Omega u_{-1}^{2}v_{-1}^{2}, \cr
h_{8}&=&g\Omega (u_{1}^{2}v_{-1}^{2}+u_{-1}^{2}v_{1}^{2}), \cr
h_{9}&=&-g\Omega u_{1}^{2}v_{1}^{2}, \cr
h_{-9}&=&-g\Omega u_{-1}^{2}v_{-1}^{2}, \cr
h_{10}&=&-2g\Omega u_{1}v_{1}u_{-1}v_{-1}.
\label{a16}
\eea\ewt
Also, in this basis the full particle number operator is given by, 
\beq
\hat{N} = \sum_{j} \hat{N}_{j}, \quad j=\pm 1
\label{a17}
\eeq
where,
\beq
\hat{N}_{j}=(u_{j}^2-v_{j}^{2})\hat{N}_{q,j}+2\Omega
v_{j}^{2}+2u_{j}v_{j}\sqrt{\Omega}(P^{\dagger}_{j}+P_{j}).
\label{a18}
\eeq
The RPA excited states are, as usual, obtained as
\beq
|\nu \rangle=Q_{\nu}^{\dagger}|RPA\rangle,
\label{a19}
\eeq
where $|RPA\rangle $ is the correlated RPA ground-state defined via the vacuum
condition 
\beq
Q_{\nu}|RPA\rangle=0.
\label{a20}
\eeq
In terms of the generators of the Hamiltonian $\hat{N}_{q,j}$, $P^{\dagger}_{j}$ and $P_{j}$, for the
most general QRPA excitation operator, which can be viewed as a Bogoliubov transformation of Fermion pair
operators~\footnote{We can not include the Hermitian pieces $\hat{N}_{q,j}$ in~(\ref{a18}), 
since this leads to non-normalizable eigenstates as in the case of Goldstone modes.}, 
we can write down the following expression 
\beq
Q_{\nu}^{\dagger}=\sum_{j=\pm 1} X_{j,\nu}\bar{P}^{\dagger}_{j} -Y_{j,\nu}\bar{P}_{j},\; \nu=1,2,
\label{a21}
\eeq
where we introduced the following notation, 
\beq
\bar{P}_{j}=\frac{P_{j}}{\sqrt{1-\frac{\langle \hat{N}_{q,j}\rangle}{\Omega}}}, \: j=\pm 1, 
\eeq
guaranteeing that the RPA excited state~(\ref{a19}) is normalized, i.e. $\langle \nu |\nu^{\prime}\rangle
=\delta_{\nu,\nu^{\prime}}$.
The RPA amplitudes $X_{j,\nu}$ and $Y_{j,\nu}$ in~(\ref{a21}) shall obey the following orthogonality 
relations,
\begin{displaymath}
\sum_{j=\pm 1}X_{j,\nu}^{2}-Y_{j,\nu}^{2}=1, \: \nu=1,2
\end{displaymath}
\begin{displaymath}
X_{-1,1}X_{-1,2}+X_{1,1}X_{1,2}-Y_{-1,1}Y_{-1,2}-Y_{1,1}Y_{1,2}=0
\end{displaymath}
\beq
X_{1,2}Y_{1,1}+X_{-1,2}Y_{-1,1}-X_{1,1}Y_{1,2}-X_{-1,1}Y_{-1,2}=0
\label{a22}
\eeq
and the closure relations,
\begin{displaymath}
\sum_{\nu=1,2}X_{j,\nu}^{2}-Y_{j,\nu}^{2}=1, \: j=\pm 1,
\end{displaymath}
\begin{displaymath}
X_{-1,1}X_{1,1}+X_{-1,2}X_{1,2}-Y_{-1,1}Y_{1,1}-Y_{-1,2}Y_{1,2}=0,
\end{displaymath}
\beq
X_{1,1}Y_{-1,1}+X_{1,2}Y_{-1,2}-X_{-1,1}Y_{1,1}-X_{-1,2}Y_{1,2}=0,
\label{c22}
\eeq  
with which one can invert relation~(\ref{a21})
\bwt\beq
\pmatrix{\bar{P}_{1} \cr \bar{P}_{-1} \cr \bar{P}_{1}^{\dagger} \cr \bar{P}_{-1}^{\dagger}}=
\pmatrix{X_{1,1}&X_{1,2}&Y_{1,1}&Y_{1,2}\cr
         X_{-1,1}&X_{-1,2}&Y_{-1,1}&Y_{-1,2}\cr
         Y_{1,1}&Y_{1,2}&X_{1,1}&X_{1,2}\cr
         Y_{-1,1}&Y_{-1,2}&X_{-1,1}&X_{-1,2}}
\pmatrix{Q_{1} \cr Q_{2} \cr Q_{1}^{\dagger} \cr Q_{2}^{\dagger}}.
\label{a23}
\eeq\ewt

In analogy to Baranger~\cite{b17} we obtain the SCQRPA equations in minimizing the following mean 
excitation energy
\beq
\Omega_{\nu}=\frac{\langle [Q_{\nu}, [H^{\prime},Q^{\dagger}_{\nu}]]\rangle}
{\langle [Q_{\nu}, Q^{\dagger}_{\nu}]\rangle}
\label{a24}
\eeq
with respect to the RPA amplitudes $X_{j,\nu}$ and $Y_{j,\nu}$. The minimization leads 
straightforwardly to the following eigenvalue problem
\bwt\beq
\pmatrix{A_{1,1}&A_{1,2}&B_{1,1}&B_{1,2} \cr
         A_{2,1}&A_{2,2}&B_{2,1}&B_{2,2} \cr
        -B_{1,1}&-B_{1,2}&-A_{1,1}&-A_{1,2} \cr
        -B_{2,1}&-B_{2,2}&-A_{2,1}&-A_{2,2}}
\pmatrix{X_{1, \nu} \cr X_{-1, \nu} \cr Y_{1, \nu} \cr Y_{-1, \nu}}=
\Omega_{\nu}\pmatrix{X_{1, \nu} \cr X_{-1, \nu} \cr Y_{1, \nu} \cr Y_{-1, \nu}}
\label{a25}
\eeq\ewt
where,
\bwt\bea
A_{1,1}&=&\langle [\bar{P}_{1}, [H^{\prime},\bar{P}^{\dagger}_{1}]] \rangle, \; 
A_{1,2}=\langle [\bar{P}_{1},[H^{\prime},\bar{P}^{\dagger}_{-1}]] \rangle \cr
A_{2,1}&=&\langle [\bar{P}_{-1},[H^{\prime},\bar{P}^{\dagger}_{1}]] \rangle,\;  
A_{2,2}=\langle [\bar{P}_{-1},[H^{\prime},\bar{P}^{\dagger}_{-1}]] \rangle\cr
B_{1,1}&=&-\langle [\bar{P}_{1},[H^{\prime},\bar{P}_{1}]] \rangle,\;  
B_{1,2}=-\langle [\bar{P}_{1},[H^{\prime},\bar{P}_{-1}]] \rangle \cr
B_{2,1}&=&-\langle [\bar{P}_{-1},[H^{\prime},\bar{P}_{1}]] \rangle,\;  
B_{2,2}=-\langle [\bar{P}_{-1},[H^{\prime},\bar{P}_{-1}]] \rangle,
\label{a26}
\eea\ewt
and the $\langle \ldots \rangle$ stands for the expectation values in the RPA vacuum defined by~(\ref{a20}).
Explicitly, the RPA matrix elements are given by
\bwt\bea
A_{1,1}&=& 2 h_{1}+h_{3}\bigg\{-\frac{2}{\Omega}\frac{\langle P^{\dagger}_{1}P_{1}\rangle}{1-\frac{\langle
\hat{N}_{q,1}\rangle}{\Omega}}+\frac{1-\frac{2\langle \hat{N}_{q,1} \rangle}{\Omega}+\frac{\langle
\hat{N}^{2}_{q,1}\rangle}{{\Omega}^{2}}}{1-\frac{\langle\hat{N}_{q,1}\rangle}{\Omega}}\bigg\}
-\frac{2}{\Omega}\frac{1}{1-\frac{\langle \hat{N}_{q,1}\rangle}{\Omega}}
\bigg\{h_{4}\langle P^{\dagger}_{-1}P_{1}\rangle+2h_{7}\langle P_{1}P_{1}\rangle+h_{8}\langle
P_{-1}P_{1}\rangle\bigg\} \cr
&+&4h_{9}\bigg\{\frac{\langle P^{\dagger}_{1}P_{1}\rangle + \langle P_{1}P^{\dagger}_{1}\rangle}
{1-\frac{\langle\hat{N}_{q,1}\rangle}{\Omega}} + \frac{\langle \hat{N}_{q,1}\rangle 
- \frac{\langle \hat{N}^{2}_{q,1}\rangle}{\Omega}}{1-\frac{\langle \hat{N}_{q,1}\rangle}{\Omega}}\bigg\}
+2h_{10}\frac{\langle \hat{N}_{q,-1}\rangle -\frac{\langle\hat{N}_{q,1}\hat{N}_{q,-1}\rangle}{\Omega}}
{1-\frac{\langle \hat{N}_{q,1}\rangle}{\Omega}} \cr
A_{1,2}&=&h_{4} \frac{1-\frac{\langle \hat{N}_{q,1}\rangle + \langle \hat{N}_{q,-1}\rangle}{\Omega}+
\frac{\langle\hat{N}_{q,1}\hat{N}_{q,-1}\rangle}{{\Omega}^{2}}}{\sqrt{\big(1-\frac{\langle
\hat{N}_{q,1}\rangle}{\Omega}\big)\big(1-\frac{\langle\hat{N}_{q,-1}\rangle}{\Omega}\big)}}
+4h_{10}\frac{\langle P_{1}P^{\dagger}_{-1}\rangle}{\sqrt{\big(1-\frac{\langle\hat{N}_{q,1}\rangle}
{\Omega}\big)\big(1-\frac{\langle\hat{N}_{q,-1}\rangle}{\Omega}\big)}} \cr
A_{2,1}&=&h_{4} \frac{1-\frac{\langle \hat{N}_{q,1}\rangle + \langle \hat{N}_{q,-1}\rangle}{\Omega}+
\frac{\langle\hat{N}_{q,1}\hat{N}_{q,-1}\rangle}{{\Omega}^{2}}}{\sqrt{\big(1-\frac{\langle\hat{N}_{q,1}\rangle}
{\Omega}\big)\big(1-\frac{\langle\hat{N}_{q,-1}\rangle}{\Omega}\big)}}
+4h_{10} \frac{\langle P_{-1}P^{\dagger}_{1}\rangle}{\sqrt{\big(1-\frac{\langle\hat{N}_{q,1}\rangle}
{\Omega}\big)\big(1-\frac{\langle\hat{N}_{q,-1}\rangle}{\Omega}\big)}} \cr
A_{2,2}&=&2h_{-1}+h_{-3}\bigg\{-\frac{2}{\Omega}\frac{\langle P^{\dagger}_{-1}P_{-1}\rangle}{1-\frac{\langle
\hat{N}_{q,-1}\rangle}{\Omega}}+\frac{1-\frac{2 \langle \hat{N}_{q,-1} \rangle}{\Omega}
+\frac{\langle \hat{N}^{2}_{q,-1}\rangle}{{\Omega}^{2}}}{1-\frac{\langle \hat{N}_{q,-1}\rangle}{\Omega}}\bigg\}
-\frac{2}{\Omega}\frac{1}{1-\frac{\langle \hat{N}_{q,-1}\rangle}{\Omega}}
\bigg\{h_{4}\langle P^{\dagger}_{1}P_{-1}\rangle+2h_{-7}\langle P_{-1}P_{-1}\rangle \cr
&+&h_{8}\langle P_{1}P_{-1}\rangle\bigg\}
+4h_{-9}\bigg\{\frac{\langle P^{\dagger}_{-1}P_{-1}\rangle + \langle P_{-1}P^{\dagger}_{-1}\rangle}
{1-\frac{\langle\hat{N}_{q,-1}\rangle}{\Omega}}+\frac{\langle \hat{N}_{q,-1}\rangle
-\frac{\langle \hat{N}^{2}_{q,-1}\rangle}{\Omega}}{1-\frac{\langle \hat{N}_{q,-1}\rangle}{\Omega}}\bigg\}
+2h_{10}\frac{\langle \hat{N}_{q,1}\rangle-\frac{\langle\hat{N}_{q,-1}\hat{N}_{q,1}\rangle}{\Omega}}
{1-\frac{\langle \hat{N}_{q,-1}\rangle}{\Omega}} \cr
B_{1,1}&=&-\frac{2}{\Omega}\frac{1}{1-\frac{\langle\hat{N}_{q,1}\rangle}{\Omega}}
\bigg\{h_{3}\langle P_{1}P_{1}\rangle+h_{4}\langle P_{-1}P_{1}\rangle
+h_{8}\langle P^{\dagger}_{-1}P_{1}\rangle\bigg\} 
+8h_{9}\frac{\langle P_{1}P_{1}\rangle}{1-\frac{\langle\hat{N}_{q,1}\rangle}{\Omega}} \cr
&+&2h_{7}\bigg\{-\frac{1}{\Omega}\frac{\langle P^{\dagger}_{1}P_{1}\rangle + \langle P_{1}P^{\dagger}_{1}\rangle}
{1-\frac{\langle\hat{N}_{q,1}\rangle}{\Omega}}
+\frac{1-\frac{2\langle \hat{N}_{q,1} \rangle}{\Omega}+\frac{\langle \hat{N}^{2}_{q,1}\rangle}
{{\Omega}^{2}}}{1-\frac{\langle\hat{N}_{q,1}\rangle}{\Omega}}\bigg\} \cr
B_{1,2}&=&h_{8}\frac{1-\frac{\langle \hat{N}_{q,1}\rangle + \langle \hat{N}_{q,-1}\rangle}{\Omega}+
\frac{\langle\hat{N}_{q,1}\hat{N}_{q,-1}\rangle}{{\Omega}^{2}}}{\sqrt{\big(1-\frac{\langle\hat{N}_{q,1}\rangle}
{\Omega}\big)\big(1-\frac{\langle\hat{N}_{q,-1}\rangle}{\Omega}\big)}}
+4h_{10}\frac{\langle P_{-1}P_{1}\rangle}{\sqrt{\big(1-\frac{\langle\hat{N}_{q,1}\rangle}
{\Omega}\big)\big(1-\frac{\langle\hat{N}_{q,-1}\rangle}{\Omega}\big)}} \cr
B_{2,1}&=&h_{8} \frac{1-\frac{\langle \hat{N}_{q,1}\rangle + \langle \hat{N}_{q,-1}\rangle}{\Omega}+
\frac{\langle\hat{N}_{q,1}\hat{N}_{q,-1}\rangle}{{\Omega}^{2}}}{\sqrt{\big(1-\frac{\langle\hat{N}_{q,1}\rangle}
{\Omega}\big)\big(1-\frac{\langle\hat{N}_{q,-1}\rangle}{\Omega}\big)}}
+4h_{10}\frac{\langle P_{1}P_{-1}\rangle}{\sqrt{\big(1-\frac{\langle\hat{N}_{q,1}\rangle}
{\Omega}\big)\big(1-\frac{\langle\hat{N}_{q,-1}\rangle}{\Omega}\big)}} \cr
B_{2,2}&=&-\frac{2}{\Omega}\frac{1}{1-\frac{\langle\hat{N}_{q,-1}\rangle}{\Omega}}
\bigg\{h_{-3}\langle P_{-1}P_{-1}\rangle + h_{4}\langle P_{1}P_{-1}\rangle + h_{8}\langle
P^{\dagger}_{1}P_{-1}\rangle \bigg\}
+8h_{-9}\frac{\langle P_{-1}P_{-1}\rangle}{1-\frac{\langle\hat{N}_{q,-1}\rangle}{\Omega}} \cr
&+&2h_{-7}\bigg\{-\frac{1}{\Omega}\frac{\langle P^{\dagger}_{-1}P_{-1}\rangle + \langle P_{-1}P^{\dagger}_{-1}\rangle}
{1-\frac{\langle\hat{N}_{q,-1}\rangle}{\Omega}}
+\frac{1-\frac{2\langle \hat{N}_{q,-1} \rangle}{\Omega}+\frac{\langle \hat{N}^{2}_{q,-1}\rangle}
{{\Omega}^{2}}}{1-\frac{\langle\hat{N}_{q,-1}\rangle}{\Omega}}\bigg\}.
\label{a27}
\eea\ewt

Using~(\ref{a23}) and the condition~(\ref{a20}) the expectation values of type 
$\langle P^{\dagger}_{j} P_{j} \rangle $, $\langle P_{j} P^{\dagger}_{j} \rangle $, 
$\langle P^{\dagger}_{j}P^{\dagger}_{j} \rangle $ and $\langle P_{j} P_{j} \rangle $ are readily 
expressed by the RPA amplitudes $X_{j,\nu}$ and $Y_{j,\nu}$ (these calculations are 
detailed in Appendix~\ref{average})
\bwt\bea
\langle P^{\dagger}_{1} P_{1} \rangle&=&\big(Y_{1,1}^{2}+Y_{1,2}^{2}\big)(1-\frac{\langle
\hat{N}_{q,1} \rangle}{\Omega}), \cr
\langle P^{\dagger}_{-1} P_{-1} \rangle&=&\big(Y_{-1,1}^{2}+Y_{-1,2}^{2}\big)(1-\frac{\langle
\hat{N}_{q,-1} \rangle}{\Omega}), \cr
\langle P_{1} P^{\dagger}_{1} \rangle&=&\big(X_{1,1}^{2}+X_{1,2}^{2}\big)(1-\frac{\langle
\hat{N}_{q,1} \rangle}{\Omega}), \cr
\langle P_{-1} P^{\dagger}_{-1} \rangle &=& \big(X_{-1,1}^{2}+X_{-1,2}^{2}\big)(1-\frac{\langle
\hat{N}_{q,-1} \rangle}{\Omega}), \cr
\langle P^{\dagger}_{1} P^{\dagger}_{1} \rangle&=&\big(X_{1,1}Y_{1,1}+X_{1,2}Y_{1,2}\big)(1-\frac{\langle
\hat{N}_{q,1} \rangle}{\Omega}), \cr
\langle P_{1} P_{1} \rangle&=&\big(X_{1,1}Y_{1,1}+X_{1,2}Y_{1,2}\big)(1-\frac{\langle
\hat{N}_{q,1} \rangle}{\Omega}), \cr
\langle P^{\dagger}_{-1} P^{\dagger}_{-1} \rangle&=&\big(X_{-1,1}Y_{-1,1}+X_{-1,2}Y_{-1,2}\big)(1-\frac{\langle
\hat{N}_{q,-1} \rangle}{\Omega}), \cr
\langle P_{-1} P_{-1} \rangle&=&(X_{-1,1}Y_{-1,1}+X_{-1,2}Y_{-1,2})(1-\frac{\langle
\hat{N}_{q,-1} \rangle}{\Omega}), \cr
\langle P_{1} P_{-1} \rangle&=&\big(X_{1,1}Y_{-1,1}+X_{1,2}Y_{-1,2}\big){\sqrt{\big(1-\frac{\langle\hat{N}_{q,1}\rangle}
{\Omega}\big)\big(1-\frac{\langle\hat{N}_{q,-1}\rangle}{\Omega}\big)}}, \cr
\langle P^{\dagger}_{1} P^{\dagger}_{-1} \rangle &=&
\big(X_{-1,1}Y_{1,1}+X_{-1,2}Y_{1,2}\big){\sqrt{\big(1-\frac{\langle\hat{N}_{q,1}\rangle}
{\Omega}\big)\big(1-\frac{\langle\hat{N}_{q,-1}\rangle}{\Omega}\big)}}, \cr
\langle P^{\dagger}_{1} P_{-1} \rangle &=&
\big(Y_{1,1}Y_{-1,1}+Y_{1,2}Y_{-1,2}\big){\sqrt{\big(1-\frac{\langle\hat{N}_{q,1}\rangle}
{\Omega}\big)\big(1-\frac{\langle\hat{N}_{q,-1}\rangle}{\Omega}\big)}}, \cr
\langle P_{1} P^{\dagger}_{-1} \rangle &=&
\big(X_{1,1}X_{-1,1}+X_{1,2}X_{-1,2}\big){\sqrt{\big(1-\frac{\langle\hat{N}_{q,1}\rangle}
{\Omega}\big)\big(1-\frac{\langle\hat{N}_{q,-1}\rangle}{\Omega}\big)}}.
\label{a28}
\eea\ewt
Before we discuss how to express the expectation values $\langle \hat{N}_{q,j}\rangle$, 
$\langle \hat{N}^{2}_{q,j}\rangle$ and $\langle \hat{N}_{q,j}\hat{N}_{q,j^{\prime}}\rangle$ as
functions of the amplitudes $X_{j,\nu}$ and $Y_{j,\nu}$, we want to give the equations for the 
determination of the Bogoliubov amplitudes $u_{j}$, $v_{j}$ of equations~(\ref{a9}) and~(\ref{a10}). 
As usual they are determined from the minimization of the ground-state energy~\cite{b11, b20}
\bea
\frac{\partial \langle H^{\prime} \rangle}{\partial u_{j}}+\frac{\partial \langle H^{\prime}\rangle}
{\partial v_{j}}\frac{\partial v_{j}}{\partial u_{j}}
& \equiv &  \langle [H^{\prime},\bar{P}^{\dagger}_{j}] \rangle  =0,\; j =\pm 1 \cr
&=& \langle [H^{\prime},Q^{\dagger}_{\nu}] \rangle  =0,\; \nu =1, 2
\label{a29}
\eea
with 
\bwt\bea
\langle H^{\prime} \rangle &=& h_{0} + h_{1} \langle \hat{N}_{q,1} \rangle 
+ h_{-1} \langle \hat{N}_{q,-1} \rangle
+h_{3} \langle P^{\dagger}_{1} P_{1} \rangle + h_{-3} \langle P^{\dagger}_{-1} P_{-1} \rangle
+h_{4}(\langle P^{\dagger}_{1} P_{-1} \rangle + \langle P^{\dagger}_{-1} P_{1} \rangle) \cr
&+&h_{7}(\langle P^{\dagger}_{1} P^{\dagger}_{1} \rangle + \langle P_{1}P_{1} \rangle)
+h_{-7}(\langle P^{\dagger}_{-1} P^{\dagger}_{-1} \rangle + \langle P_{-1}P_{-1} \rangle)
+h_{8}(\langle P^{\dagger}_{1} P^{\dagger}_{-1} \rangle + \langle P_{-1}P_{1} \rangle) \cr
&+&h_{9}\langle \hat{N}_{q,1}^{2} \rangle + h_{-9}\langle \hat{N}_{q,-1}^{2} \rangle
+h_{10} \langle \hat{N}_{q,1}\hat{N}_{q,-1}\rangle.
\label{a30}
\eea\ewt
It should be mentioned that when~(\ref{a29}) is evaluated with a BCS ground-state
then~(\ref{a29}) leads to the usual BCS equations. However, here we use the
correlated RPA ground-state and then the mean field equations couple back to the
RPA amplitudes $X_{j,\nu}$ and $Y_{j,\nu}$.
Explicitly these equations lead to 
\beq
2\xi_{j}u_{j}v_{j}+\Delta_{j}(v^{2}_{j}-u^{2}_{j})=0,\; j=\pm 1
\label{a31}
\eeq
which together with~(\ref{a11}) can be written as 
\begin{equation}
\pmatrix{ \xi_{j} & \Delta_{j} \cr \Delta_{j} & -\xi_{j} \cr }
\pmatrix{ u_{j} \cr  v_{j} \cr } = E_{j} \pmatrix{ u_{j} \cr v_{j} \cr }, \;
E_{j}=\sqrt{\xi_{j}^{2}+\Delta_{j}^{2}}
\label{a32}
\end{equation}
with the standard solution
\bea
u^{2}_{j}=\frac{1}{2}(1+\frac{\xi_{j}}
{\sqrt{{\xi_{j}^{2}+\Delta_{j}^{2}}}}), \cr
v^{2}_{j}=\frac{1}{2}(1-\frac{\xi_{j}}
{\sqrt{{\xi_{j}^{2}+\Delta_{j}^{2}}}})
\label{a33}
\eea
from where follows the gap equation,
\beq
\Delta_{i}=\sum_{j}\tilde{g}_{i j}u_{j}v_{j}=\frac{1}{2}\sum_{j}\tilde{g}_{i j}\frac{\Delta_{j}}
{\sqrt{\xi_{j}^{2}+\Delta^{2}_{j}}},\; i, j = \pm 1
\label{a34}
\eeq
where the renormalized single-particle energies are
\bwt\beq
\xi_{j}=(j\frac{\epsilon}{2}-g v^{2}_{j})
+\frac{g}{1-\frac{\langle \hat{N}_{q,j}\rangle}{\Omega}}(u^{2}_{-j}-v^{2}_{-j})
\big(\langle P^{\dagger}_{j} P^{\dagger}_{-j} \rangle 
+ \langle P^{\dagger}_{j} P_{-j} \rangle\big) - \mu,\; j = \pm 1
\label{a35}
\eeq\ewt
and the renormalized interaction is given by
\beq
\tilde{g}=\pmatrix{\tilde{g}_{1,1} & \tilde{g}_{1,-1} \cr
                   \tilde{g}_{-1,1}& \tilde{g}_{-1,-1} },
\label{a36}
\eeq
with,
\bwt\bea
\tilde{g}_{1,1} &=& g\Omega -\frac{g}{1-\frac{\langle \hat{N}_{q,1}\rangle}{\Omega}}
\bigg\{2\big(\langle P^{\dagger}_{1}
P^{\dagger}_{1} \rangle +\langle P^{\dagger}_{1}
P_{1} \rangle\big) +\langle \hat{N}_{q,1} \rangle 
-\frac{\langle \hat{N}_{q,1}^{2} \rangle}{\Omega}\bigg\} \cr
\tilde{g}_{1,-1}&=& g\Omega-g \frac{\langle \hat{N}_{q,-1} \rangle
-\frac{\langle \hat{N}_{q,1} \hat{N}_{q,-1} \rangle}{\Omega}}
{1-\frac{\langle \hat{N}_{q,1} \rangle}{\Omega}} \cr
\tilde{g}_{-1,1}&=&g\Omega -g\frac{\langle \hat{N}_{q,1} \rangle
-\frac{\langle \hat{N}_{q,1} \hat{N}_{q,-1} \rangle}{\Omega}}
{1-\frac{\langle \hat{N}_{q,-1} \rangle}{\Omega}} \cr
\tilde{g}_{-1,-1} &=& g\Omega-\frac{g}{1-\frac{\langle \hat{N}_{q,-1}\rangle}{\Omega}}
\bigg\{2\big(\langle P^{\dagger}_{-1}
P^{\dagger}_{-1} \rangle +\langle P^{\dagger}_{-1}
P_{-1} \rangle\big) +\langle \hat{N}_{q,-1} \rangle 
-\frac{\langle \hat{N}_{q,-1}^{2} \rangle}{\Omega}\bigg\}.
\label{a37}
\eea\ewt
We see that the mean field equations have exactly the same mathematical structure as in the BCS case, 
however, with renormalized vertices and single-particle energies involving the RPA amplitudes. 
We, therefore, explicitly see that the mean field equations are coupled to the quantum fluctuations.

Let us now come to the elaboration of the quasi-particles occupation numbers and their variances. The
determination of those quantities is one of the difficulties in the SCQRPA approach~\cite{b7, b11, b20}. 
However, this problem has found an elegant solution in the early works of~\cite{b21} (see also~\cite{b33}). 
In the same way, we derived expressions of the quasi-particles occupation numbers and their variances as
expansions in the operators $P^{\dagger}_{j}$ and $P_{j}$ up to any order in a systematic way. The detailed
derivation is given in the Appendix~\ref{Number}. We here present a different method which shows some 
interesting aspects and will lead to the same result. Using the bosonic representation of the quasi-spin 
operators of our model, we can write 
\bea
\hat{N}_{q,j}&=& 2 B^{\dagger}_{j}B_{j}, \cr
P^{\dagger}_{j}&=&B^{\dagger}_{j}\bigg(1-\frac{1}{\Omega}B^{\dagger}_{j}B_{j}\bigg)^{\frac{1}{2}},
\cr
P_{j}&=&(P^{\dagger}_{j})^{\dagger}=\bigg(1-\frac{1}{\Omega}B^{\dagger}_{j}B_{j}\bigg)^{\frac{1}{2}}
B_{j}
\label{a38}
\eea
where, one can show that these operators in this representation always obey to
the commutation rules of angular momentum~(\ref{a39}).
We also can invert this relation, and we obtain
\bea
B^{\dagger}_{j}&=&P^{\dagger}_{j}\bigg(1-\frac{1}{\Omega}B^{\dagger}_{j}B_{j}\bigg)^{-\frac{1}
{2}}, \cr
B_{j}&=&\bigg(1-\frac{1}{\Omega}B^{\dagger}_{j}B_{j}\bigg)^{-\frac{1}{2}}P_{j}.
\label{a40}
\eea
With~(\ref{a40}) $\hat{N}_{q,j}$ can be expressed as
\bea
\hat{N}_{q,j}&=& 2 B^{\dagger}_{j}B_{j}, \cr
&=& 2
P^{\dagger}_{j}\bigg(1-\frac{1}{\Omega}B^{\dagger}_{j}B_{j}\bigg)^{-1}P_{j}, \cr
&=& 2 P^{\dagger}_{j}\bigg(1-\frac{1}{2\Omega}\hat{N}_{q,j}\bigg)^{-1}P_{j}.
\label{a41}
\eea
Therefore, we obtained a recursive relation for $\hat{N}_{q,j}$, and with it we can derive
an expansion for $\hat{N}_{q,j}$. By successive replacement of $\hat{N}_{q,j}$ in 
the rhs of~(\ref{a41}), one finds the following expansion,
\bea
\hat{N}_{q,j} &=& 2
P^{\dagger}_{j}\bigg(1-\frac{1}{\Omega}P^{\dagger}_{j}P_{j}\bigg)^{-1}P_{j}, \cr
&=&2 P^{\dagger}_{j}P_{j} +
\frac{2}{\Omega}{P^{\dagger}_{j}}^{2}\sum_{n=0}^{\infty}
\bigg(\frac{P_{j}P^{\dagger}_{j}}{\Omega}\bigg)^{n}P_{j}^{2}, \cr
&=&2 P^{\dagger}_{j}P_{j} +
\frac{2}{\Omega}{P^{\dagger}_{j}}^{2}\sum_{n=0}^{\infty}
\bigg(\frac{\Omega P^{\dagger}_{j}P_{j}-\hat{N}_{q,j}
+\Omega}{\Omega^{2}}\bigg)^{n}P_{j}^{2}, \cr
&= &2 P^{\dagger}_{j}P_{j} +
\frac{2}{\Omega}{P^{\dagger}_{j}}^{2}\sum_{n=0}^{\infty}
\bigg(\frac{1}{\Omega}\bigg)^{n}P_{j}^{2}+ \ldots, \cr
&= & 2 P^{\dagger}_{j}P_{j}+\frac{2}{\Omega -1}{P^{\dagger}_{j}}^{2}P_{j}^{2}+ \ldots.
\label{a44}
\eea 
It should be noted that the first term in~(\ref{a44}) becomes already exact for $J=1/2$ and including
the second term it is also exact for $J=3/2$, etc. 

For $\hat{N}^{2}_{q,j}$, we can use the Casimir relation,
\beq
\Omega P^{\dagger}_{j}P_{j} + \frac{\hat{N}^{2}_{q,j}}{4}-\frac{\Omega
+1}{2}\hat{N}_{q,j}=0.
\label{casimir}
\eeq  
It is equivalent to use the expansion of $\hat{N}^{2}_{q,j}$ obtained as 
the square of $\hat{N}_{q,j}$,
\bea
\hat{N}_{q,j}^{2} = 4
P^{\dagger}_{j}P_{j}+\frac{4(\Omega+1)}{(\Omega-1)}{P^{\dagger}_{j}}^{2}
P_{j}^{2}+\ldots.
\label{a45}
\eea 
In the same way, we use~(\ref{a44}) to obtain an expansion for $\hat{N}_{q,1} \hat{N}_{q,-1}$, but it is 
sufficient to use the term of the first order of this expansion, to obtain
\bea
\hat{N}_{q,1} \hat{N}_{q,-1} = 4 P^{\dagger}_{1}P_{1}P^{\dagger}_{-1}P_{-1}+\ldots
\label{a46}
\eea
In principle the expansion~(\ref{a44}) can be pushed to higher order, however, it quickly becomes 
quite cumbersome and in practice we always will stop at second order. In any case the expansion is 
finite with maximal $J+1/2$ terms. It is natural that such an expansion exists since there is a duality
between the pair of operators $B^{\dagger}_{j}, B_{j} \Leftrightarrow P^{\dagger}_{j}, P_{j}$. There is
the choice either to bosonize the problem then everything is expressed in terms of $B^{\dagger}_{j}$ 
and $B_{j}$ operators. Or one stays with the Fermion pair operators and everything is expressed in terms
of $P^{\dagger}_{j}$ and $P_{j}$. In~\cite{b23} the former route was chosen, here we choose the latter one. 
One should mention that a truncation of the series~(\ref{a44}) also entails some violation of the Pauli 
principle but one may notice that the series is very fast converging and that already the lowest order 
correctly contains two limits : $J=1/2$ as already mentioned and $J \longrightarrow \infty$, since then 
$P^{\dagger}_{j}\longrightarrow B^{\dagger}_{j}$ and the lowest order is also correct see~(\ref{a38}). 
With these remarks in mind we go ahead. By the inversion of the QRPA excitation operator 
$Q^{\dagger}_{\nu}$, the expectation values of these expressions are immediately given in terms of the RPA
amplitudes $X_{j,\nu}$ and $Y_{j,\nu}$, as one can see in appendix \ref{average} where we give some 
details concerning the calculation of expectations values of these expressions in the RPA ground-state.

Our system of SCQRPA equations is now fully closed and we can proceed to its solution. Before let us, 
however, shortly come back to the limit of standard QRPA. This we will do for the symmetric case i.e. 
$N=2\Omega$. This case is obtained in evaluating all expectation values in all interaction kernels with
the BCS ground-state or else putting $Y_{j,\nu}=0$ and $\sum_{\nu}X^{2}_{j,\nu}=1$ for $j=\pm 1$. 
The matrix elements are then
\bea
A_{1,1}&=&A_{2,2}=g\Omega-\frac{{\Delta}^{2}}{2g{\Omega}^{2}}+\frac{{\Delta}^{2}}{2g\Omega}, \cr
A_{1,2}&=&A_{2,1}=-\frac{{\Delta}^{2}}{2g\Omega}, \cr
B_{1,1}&=&B_{2,2}=-\frac{{\Delta}^{2}}{2g\Omega^{2}}+\frac{{\Delta}^{2}}{2g\Omega}, \cr
B_{1,2}&=&B_{2,1}=g\Omega-\frac{{\Delta}^{2}}{2g\Omega}
\label{a47}
\eea
where, the gap equation in the BCS theory leads to the solution in the symmetric case
\beq
\Delta=\sqrt{g^{2}{\Omega}^{2}-\frac{{\xi}^{2}}{4}},
\label{a48}
\eeq
together with,
\bea
u^{2}_{1}&=&v^{2}_{-1}=\frac{1}{2}\bigg(1+\frac{\xi}{2 g \Omega}\bigg), \cr
v^{2}_{1}&=&u^{2}_{-1}=\frac{1}{2}\bigg(1-\frac{\xi}{2 g \Omega}\bigg), \cr
\mu&=&-\frac{g}{2}
\label{a49}
\eea
where $\xi$ is defined as $\xi=2\epsilon\Omega/(2\Omega-1)$.
For the positive eigenvalues of the RPA matrix, we obtain
\beas
\Omega^{QRPA}_{1}&=&0, \slabel{w1}\\
\Omega^{QRPA}_{2}&=&\sqrt{4{\Delta}^{2}-\frac{2{\Delta}^{2}}{\Omega}}.\slabel{w2}
\label{a50}
\eeas
As usual the other two eigenvalues are $-\Omega^{QRPA}_{\nu}$ with $\nu =1, 2$. These results are well
known~\cite{c24, b24}. We have repeated them here for completeness and stressing the point that in QRPA,
because of the spontaneously broken particle number symmetry, one obtains a Goldstone mode
$\Omega^{QRPA}_{1}=0$. We again would like, to stress the point that this is the case only if we
evaluate~(\ref{a47}) with the solution $u_{j}$, $v_{j}$ given by the mean field equations
(\ref{a31}) which for $\sum_{\nu}X^{2}_{j,\nu}=1$, $Y_{j,\nu}=0$ reduce to the usual BCS equations.
We explicitly showed it here for the symmetric case but the same scenario holds true for cases away from
half filling.
\section{Results and Discussions}
\label{results}
We first recall that the phase transition point in BCS theory for the two level pairing model is produced 
at $g_{c}=\epsilon/(2\Omega -1)$, where $\epsilon$ is the single-particle energy splitting and $\Omega$ is
the pair degeneracy of each level. In the following, the graphs are plotted, as usual, as function of the 
variable $ V=g\Omega/2\epsilon $, and refer to the case with level spin $J=11/2$, i.e. $\Omega = 6$ and 
single-particle energy $\epsilon = 2$ (in arbitrary units). This latter value for $J$ has been chosen for
easier comparison with the results of~\cite{b23} which will be given in section~\ref{compar}.  

\begin{figure}                       
\resizebox{0.5\columnwidth}{!}{\includegraphics{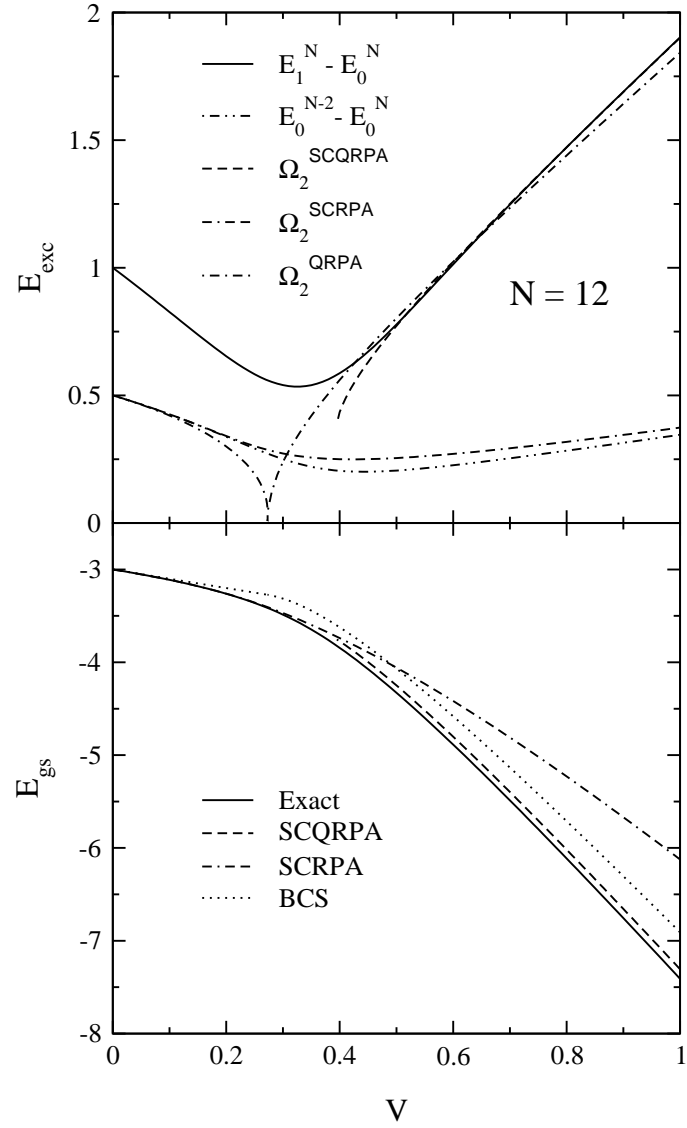}}
\caption{Ground-state energy $E_{gs}$ and excitation energy of the first $0^{+}$ state $E_{exc}$ as a function 
of the variable $V=g\Omega/2\epsilon$ described in the text and for particle number $N=12$ 
(energies are divided by $2\epsilon$). The spin of the levels is $J=11/2$. The results refer to exact 
calculations (solid line and double-dot dashed line), BCS (dotted line), RPA and QRPA (dot-dashed line), 
SCQRPA (dashed line) and SCRPA (double-dash dotted line). (Note that SCRPA and SCQRPA solutions co-exist
over a wide range of $V$-values).}
\label{full1}
\end{figure}

Let us first discuss the case with  $N=12$, i.e. the lower level is filled in the absence of correlations. 
We call this the half-filled or symmetric case. In Fig.\ref{full1} we show in the upper
panel the excitation energies. Let us consider the well known scenario of the
standard RPA. Before the phase transition to the superfluid phase we work with the
unconstrained Hamiltonian. One obtains two eigenvalues with the interpretation of
differences of ground-state energies, differing by two units in mass
$2\mu^{\pm}=\pm(E^{N\pm 2}_{0}-E^{N}_{0})$. They are evidently related to the
chemical potential and in standard particle-particle RPA (pp-RPA) they are given by
\beas
\Omega_{a} &=& 2\mu^{+} = -g+\sqrt{g+\epsilon}\sqrt{\epsilon+g(1-2\Omega)}, \slabel{mu+}\\
\Omega_{r} &=& -2\mu^{-} = g+\sqrt{g+\epsilon}\sqrt{\epsilon+g(1-2\Omega)}, \slabel{mu-}
\label{a51}
\eeas
where $\Omega_{a}$ and $\Omega_{r}$ correspond to the addition and removal phonons of the pp-RPA, 
respectively. In Fig.\ref{full1} the case $-2 \mu^{-} $ is shown and we will discuss the case $2\mu^{+}$
separately below in Fig.\ref{eigen12}. We see on the graph the usual result, namely that 
$-2 \mu^{-} $ drops to zero at the phase transition point (strictly speaking only in the large $\Omega $ 
limit). After the phase transition point we work with the constrained Hamiltonian~(\ref{a14}) in the BCS 
quasi-particle representation. The QRPA eigenvalue~(\ref{w2}) is also shown in Fig.\ref{full1}. 
The Goldstone mode~(\ref{w1}) at zero energy corresponds to a rotation in gauge space whereas the second 
eigenvalue corresponds to the "$\beta-$vibration"  of the nucleus with $N$ particles~\cite{b18}. This 
difference in interpretation is also well born out in the SCQRPA in comparison with the exact solution. 
We see that in the transition region SCRPA shows a tremendous improvement over RPA and that SCRPA follows 
the exact value of $-2\mu^{-}$ even far beyond the phase transition point (as defined by BCS theory) where 
no RPA solution exists. It is also to be noticed that the sharp phase transition seen in RPA-QRPA is an 
artifact of the theory and that in reality the phase transition is completely washed out due to the 
finiteness of the system. The fact that the "spherical" SCRPA solution co-exists with the "deformed"
SCQRPA solution over a wide parameter range representing different energy states of the system is a quite 
unique situation. In all other model cases where we have investigated the "spherical-deformed" transition 
the "spherical" solution ceased to converge numerically~\cite{b25} beyond a certain critical coupling. 
This, however, is no proof that the "spherical" solution does not also exist far in the deformed region 
representing physical states. It may be that in those works simply the method for the numerical solution 
was not sophisticated enough. This is a point to be investigated in the future. In the superfluid 
(deformed) region SCQRPA still is superior to QRPA but the improvement is less spectacular. This 
stems from the fact that the transformation to BCS quasi-particles effectively accounts already for some 
supplementary correlations in QRPA and thus the differences with exact and SCQRPA solutions become less 
important than in the non superfluid regime. A feature which is to be remarked in Fig.\ref{full1} is the 
fact that SCRPA and SCQRPA do not smoothly match in the transition region whereas RPA and QRPA have a 
certain continuity at the transition point. However, we see that SCRPA and SCQRPA describe two physically 
very distinct states which do not have any contact in the exact case neither and therefore it is not 
astonishing that SCRPA and SCQRPA do not join. This mismatch has as a consequence that there also exists 
a rupture in the ground-state energy as a function of interaction as is seen in the lower panel of
Fig.\ref{full1}. Again SCQRPA results improve strongly over BCS ground-state energies in the deformed 
region.

\begin{figure}                       
\resizebox{0.65 \columnwidth}{!}{\includegraphics{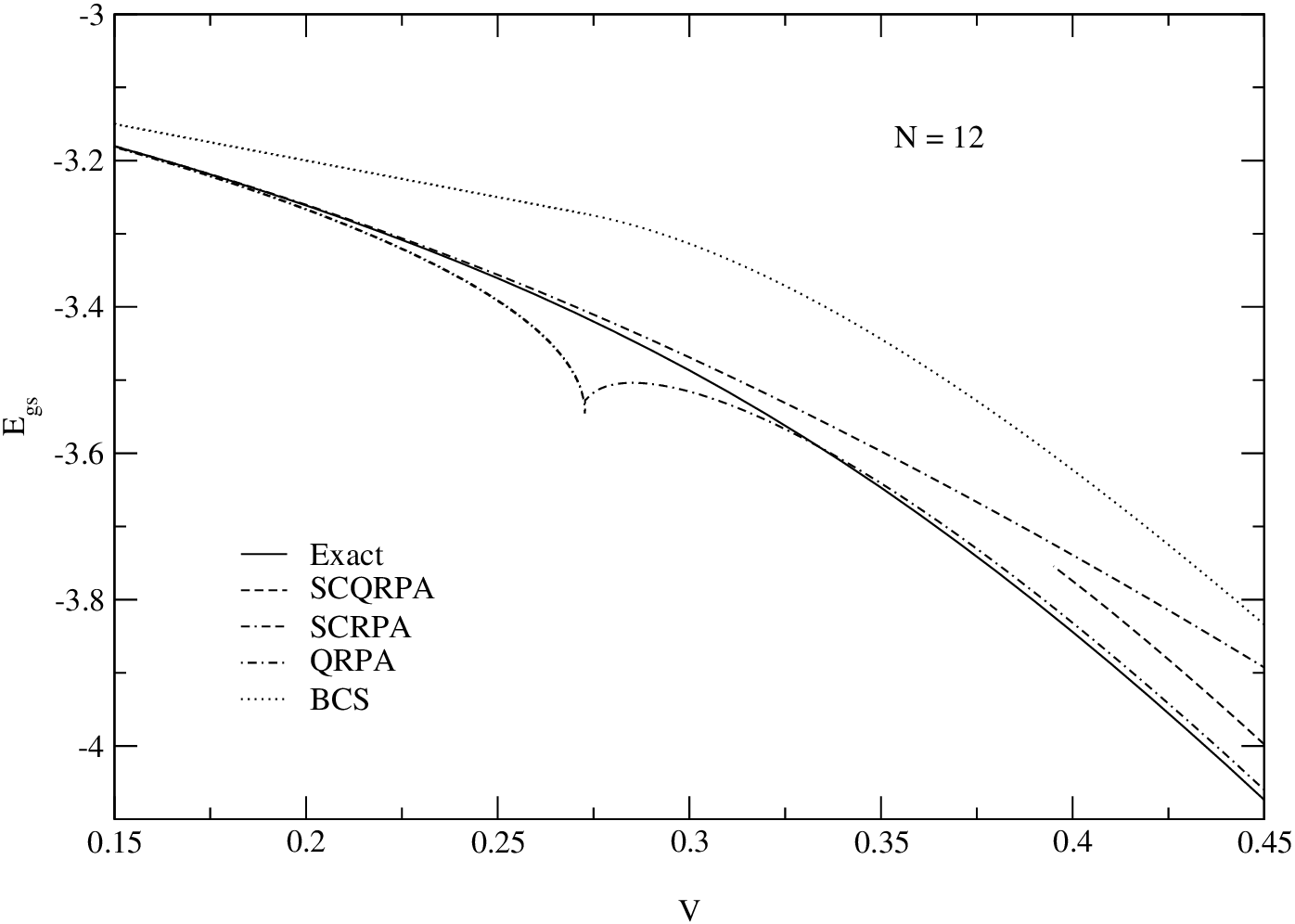}}
\caption{Zoom on the ground-state energy $E_{gs}$ as a function of the variable $V=g\Omega/2\epsilon$ 
described in the text and for particle number $N=12$ (energies are divided by $2\epsilon$). The spin 
of the levels is $J=11/2$. The results refer to exact calculations (solid line), BCS (dotted line), 
RPA and QRPA (dot-dashed line), SCQRPA (dashed line) and SCRPA (double-dash dotted line).}
\label{full2}
\end{figure}

So far we have omitted the discussion of two items of the case considered in Fig.\ref{full1} which
are slightly subtle. The first is the fact that the QRPA shows two eigenvalues : the
"$\beta-$vibration" and the Goldstone mode at zero energy ("the pair rotation mode"), whereas we
have not shown the corresponding low energy mode of SCQRPA. We will below devote an extra paragraph to
this issue. The second point is that we have not shown in Fig.\ref{full1} the QRPA values for the
ground-state energies. We show this separately in an enlarged scale around the transition point in
Fig.\ref{full2}. We there see that QRPA overbinds in the transition region but that further to the
right of the transition region QRPA values are closer to the exact solution than the ones from SCQRPA.
This is a paradoxical result which systematically repeats itself for all other configurations we
will consider below. However, the seemingly "better agreement" is an artifact of the QRPA which has
already been encountered in others cases~\cite{b25}. We want to argue as follows : SCQRPA is in itself 
a well defined theory, resulting from the variational principle~(\ref{a24}) for two body correlation 
functions. One also can consider it as a HFB approach for Fermion pairs. The Pauli principle is respected 
in an optimal way, since at no point a bosonization of Fermion pair operators is introduced and the Pauli 
principle is only violated in the truncation of~(\ref{a44}) which is a very fast converging series. 
However, any approximation to the full SCQRPA scheme necessarily diminishes the respect of the Pauli 
principle what simulates more correlations than there should be. Since for the present model case the 
SCQRPA ground-state energy is systematically above the exact one (under binding), it may happen that, when
the Pauli principle constraint is released in going from SCQRPA to QRPA, the corresponding gain in energy
is such that, accidentally, the QRPA ground-state energy practically coincides with the exact values
over a wide rang of parameters. We think that this is what happens in this model not only for
the configuration in Fig.\ref{full1} but systematically for all types of degeneracies and all fillings. 
We will not discuss this issue for the other cases any more in this work. We again should mention that we 
have found such fortuitous coincidences already in other works~\cite{b25}. However, in more realistic 
cases ones usually finds that the standard RPA strongly overbinds with respect to the exact 
values~\cite{b26}.

\begin{figure}                       
\resizebox{0.5\columnwidth}{!}{\includegraphics{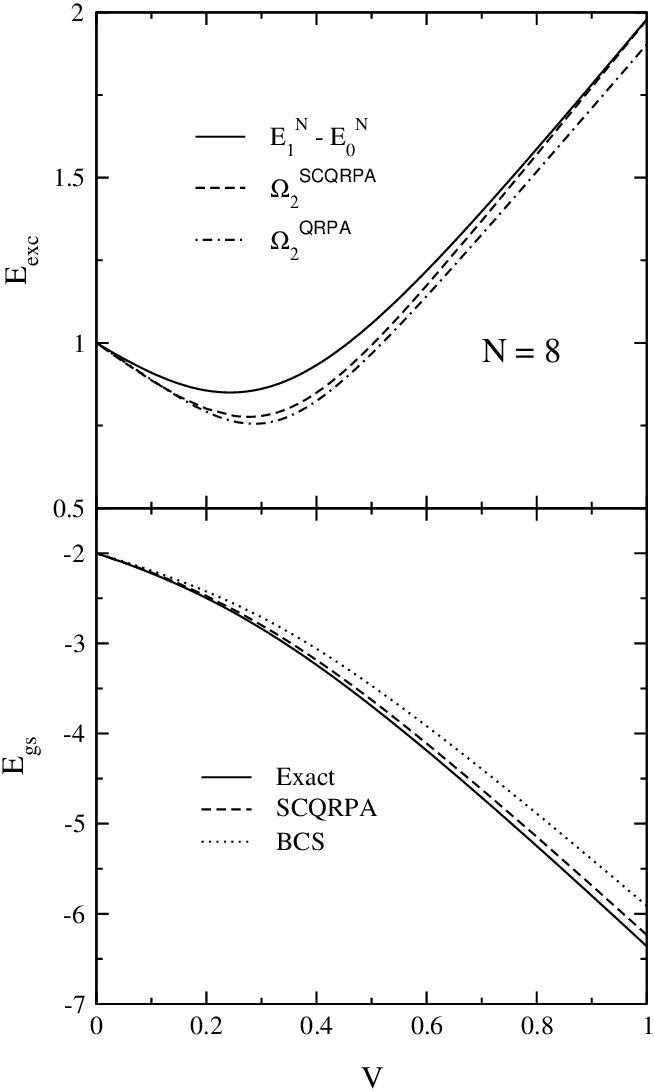}}
\caption{As in Fig.\ref{full1} but for $N=8$.}
\label{full3}
\end{figure}

\begin{figure}                       
\resizebox{0.5\columnwidth}{!}{\includegraphics{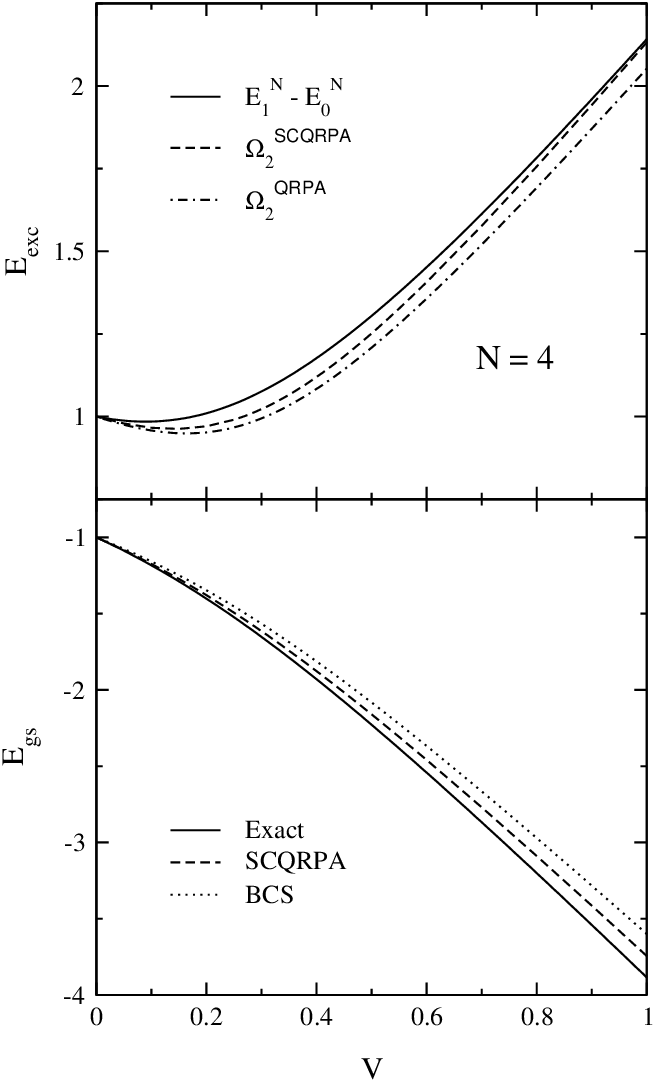}}
\caption{As in Fig.\ref{full1} but for $N=4$.}
\label{full4}
\end{figure}

\begin{figure}                       
\resizebox{0.5\columnwidth}{!}{\includegraphics{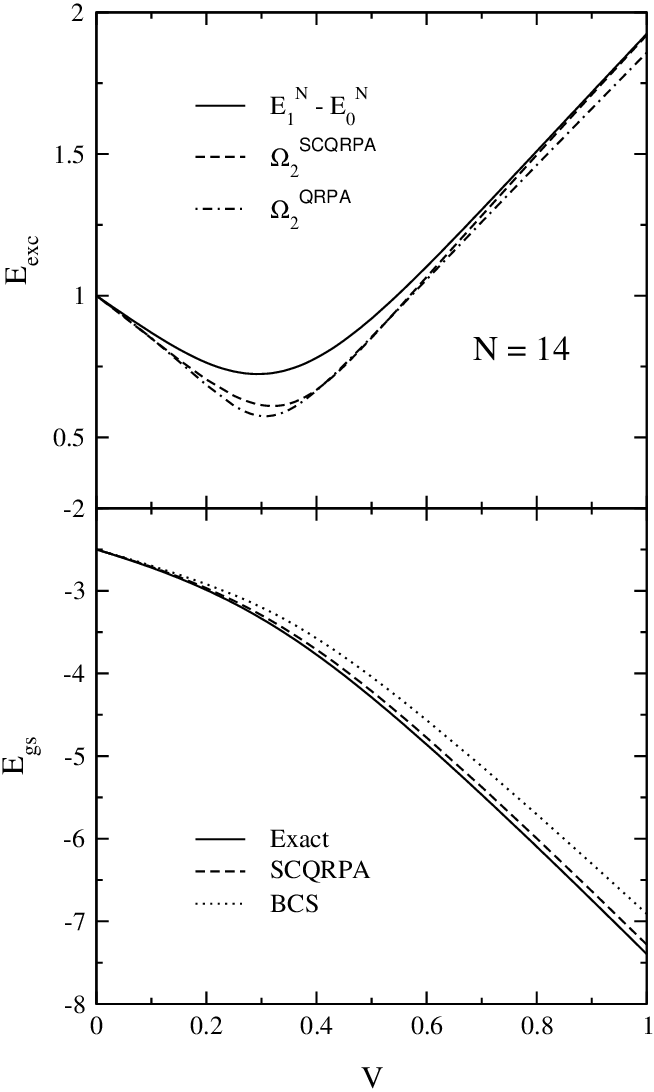}}
\caption{As in Fig.\ref{full1} but for $N=14$.}
\label{full5}
\end{figure}

Let us now discuss situations where either the lower or upper levels are only partially filled. Like in
the one level pairing case these configurations always show a non trivial BCS solution, i.e. they are
always in the superfluid regime independent of $V$. Let us look at Fig.\ref{full3} with $J=11/2$ and
$N=8$ that is the lower level partially filled for $V=0$. In the upper panel the high lying eigenvalue 
of the SCQRPA equations is shown against the exact value. We see that there is some improvement
of SCQRPA with respect to QRPA but it is not spectacular. It is similar to the case of Fig.\ref{full1}
where in the superfluid region the improvement, for reasons already explained above, is modest.
For the ground-state energy there is quite strong improvement over BCS theory. The QRPA result is not
shown but the situation is the same as already explained above. The cases $J=11/2$, $N=4$ and $N=14$ shown
in Fig.\ref{full4}-\ref{full5} are qualitatively similar. 

\begin{figure}                       
\resizebox{0.65 \columnwidth}{!}{\includegraphics{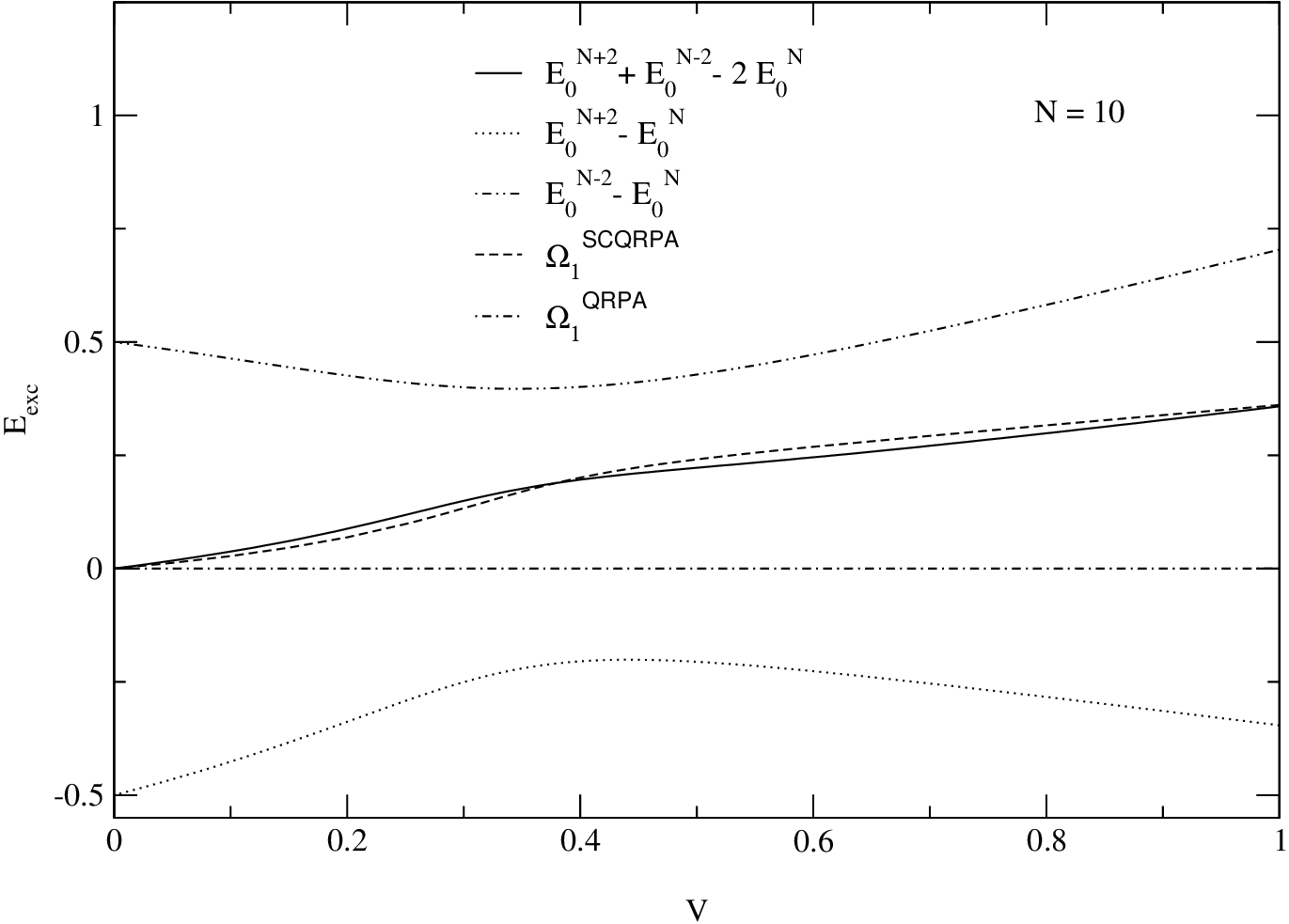}}
\caption{Excitation energy of the soft (spurious) mode (energies are divided by $2 \epsilon$) 
as a function of the variable $V=g\Omega/2\epsilon$ described in the text and for particle 
number $N=10$. The spin of the levels is $J=11/2$. The results refer to exact calculations 
(solid line, dotted line and double-dot dashed line), QRPA (dot-dashed line), and SCQRPA (dashed line).}
\label{eigen10}
\end{figure}

\begin{figure}                       
\resizebox{0.65 \columnwidth}{!}{\includegraphics{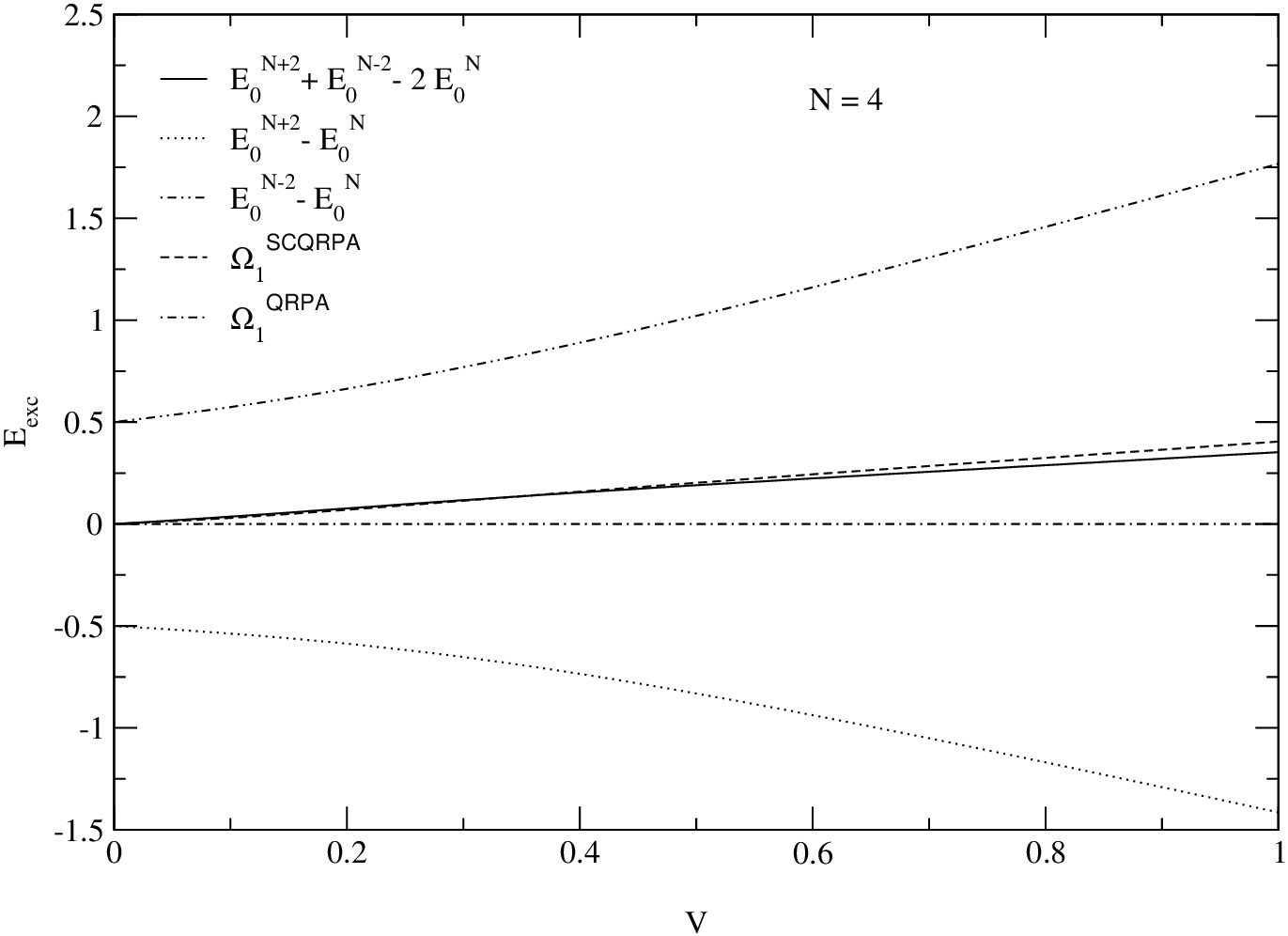}}
\caption{As in Fig.\ref{eigen10} but for $N=4$.}
\label{eigen4}
\end{figure}

Let us now come to the low lying eigenvalue of SCQRPA which in QRPA corresponds to the zero energy
eigenvalue (Goldstone or spurious mode). In Fig.\ref{eigen10} we show the low lying eigenvalue for the
case $J=11/2$ and $N=10$. We see that this eigenvalue follows very precisely the difference
$2(\mu^{+}-\mu^{-})=E^{N+2}_{0}+E^{N-2}_{0}-2E^{N}_{0}$ of the two chemical potentials
$2\mu^{+}=E^{N+2}_{0}-E^{N}_{0}$ and $2\mu^{-}=E^{N}_{0}-E^{N-2}_{0}$ as obtained
from the exact calculation. This identification makes indeed sense : since we are in the symmetry broken phase
the SCQRPA system can not distinguish between $N\pm 2$ states. For large $N$ both $2\mu^{+}$ and
$2\mu^{-}$ tend individually to Goldstone modes but for finite $N$ it definitely is reasonable to define
the difference between $2\mu^{+}$ and $2\mu^{-}$ as the low lying excitation and it is this combination
which shows up as low lying mode in the SCQRPA calculation. This is confirmed in looking at other
configurations : in Fig.\ref{eigen4} we show the case $J=11/2$, $N=4$ and in fact we find analogous
scenarios for all configurations we investigated, besides one : this is the symmetric case with
$J=11/2$, $N=12$. In the Fig.\ref{eigen12} we see that the picture is slightly different from the rest of
the cases. This stems from the fact that in the symmetric case we have a transition from the
superfluid to the non superfluid regime which is absent in the other partially filled cases. We also see that
the values for $2\mu^{+}$ and $2\mu^{-}$ are very asymmetric, $2\mu^{+}$ apparently taking the role of the
Goldstone mode alone. Also the agreement of the low lying SCQRPA solution $\Omega_{1}$ is slightly less good
than in all other cases.

Let us also add some remarks why in SCQRPA there is, contrary to QRPA, no exact Goldstone mode at
zero energy. This is relatively easy to understand : in quasi-particle representation the number
operator is given by~(\ref{a17}),~(\ref{a18}). One can check that in QRPA the terms
$\alpha^{\dagger}\alpha$, if they were included, completely decouple of the QRPA equations. Therefore, in QRPA
it is as if one had used the full particle number operator and therefore a particular solution of the QRPA
equations is $Q^{\dagger}\equiv\hat{N}$ and with $[H, \hat{N}]=0$ we get the zero eigenvalue in the Equation of
Motion (EOM) approach. 
This argumentation is no longer true in SCQRPA where the terms $\hat{N}_{q,j}$~(\ref{a18}) of the number operator
contribute in principle to SCQRPA. However, we can not include them in the RPA operator because these 
are hermitian pieces leading to non-normalizable eigenstates. Therefore $Q^{\dagger}=\hat{N}$ as a 
particular solution only holds in QRPA but not in other cases such as SCQRPA. However as a benefit, we see
in the preceding figures that we can identify the finite value of $\Omega_{1}$ with a particular 
rotational frequency in gauge space of the exact solution of the problem. On the other hand in realistic 
situation one can include in the RPA operator terms of the form $\alpha^{\dagger}_{k}\alpha_{k^{\prime}}$
for $k \neq k^{\prime}$~\cite{b12}. Only the hermitian operators $\alpha^{\dagger}_{k}\alpha_{k}$ have to be 
excluded for the reason already mentioned. These components correspond in an infinite system to momentum transfer 
zero and they are thus of zero measure. Therefore in an infinite system we have again full restoration of
symmetry.
 
\begin{figure}                       
\resizebox{0.5\columnwidth}{!}{\includegraphics{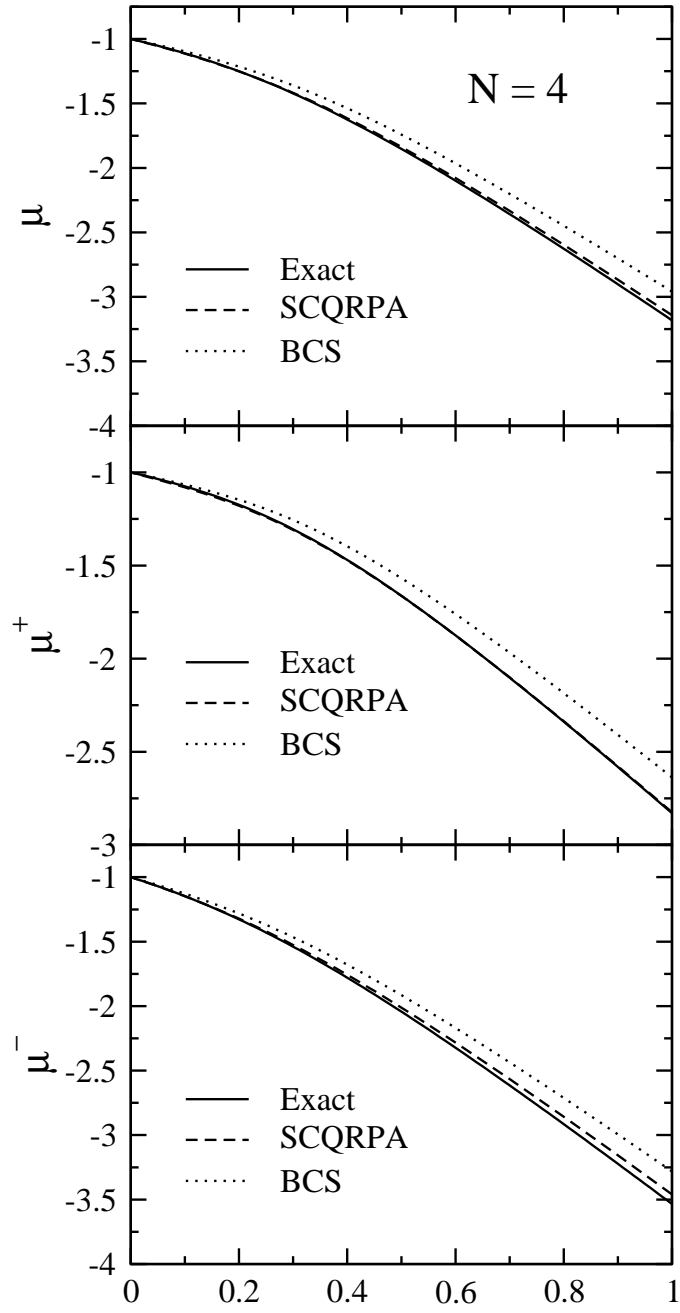}}
\caption{Comparison between SCQRPA, BCS and exact results for the chemical potentials
$\mu=\frac{1}{2}(\mu^{+}+\mu{-})$ and $\mu^{\pm}=\pm\frac{1}{2}(E^{N\pm 2}_{0}-E^{N}_{0})$,
for particle number $N=4$. The spin of the levels is $J=11/2$. The results refer to exact 
calculations (solid line), SCQRPA (dashed line) and BCS (dotted line).}
\label{pot1}
\end{figure}

\begin{figure}                       
\resizebox{0.5\columnwidth}{!}{\includegraphics{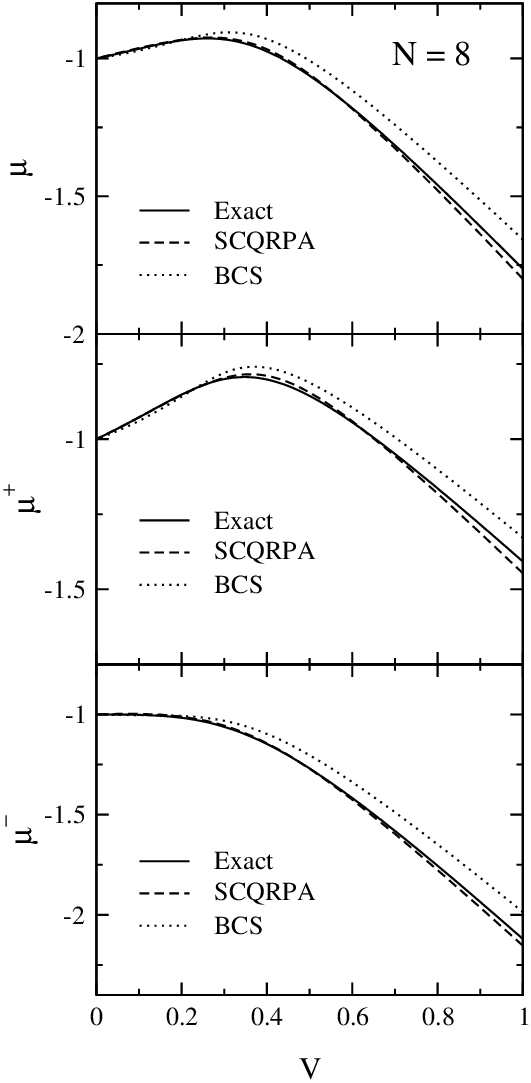}}
\caption{As in Fig.\ref{pot1} but for $N=8$.}
\label{pot2}
\end{figure}

\begin{figure}                       
\resizebox{0.5\columnwidth}{!}{\includegraphics{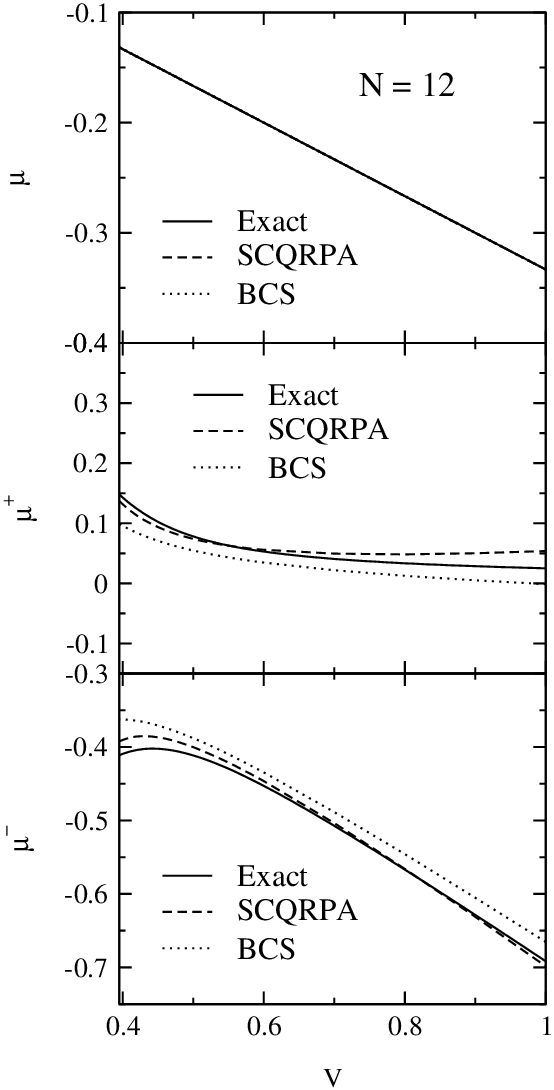}}
\caption{As in Fig.\ref{pot1} but for $N=12$.}
\label{pot3}
\end{figure}

Other quantities which are interesting to be calculated within the SCQRPA formalism are the chemical
potentials directly from differences of ground-state energies. For example in
Fig.\ref{pot1}-\ref{pot2} we show $\mu^{\pm}=\pm\frac{1}{2}(E^{N\pm 2}_{0}-E^{N}_{0})$
where the individual ground-state energies are obtained directly from separate SCQRPA calculations.
We see for $J=11/2$ and $N=4\;\text{and}\;8$ that the agreement between SCQRPA results and exact values is
excellent and in any case a strong improvement over BCS theory can be noticed. The same is true for
the chemical potential $\mu$ as obtained from $\mu=\frac{1}{2}(\mu^{+}+\mu^{-})$ in the exact calculation.
This latter which is an average chemical potential should be identified with the Lagrange multiplier $\mu$
used for restoring the symmetry of the good particle number~(\ref{a8}) in BCS and SCQRPA. This identification
is shown in each upper panel in Fig.\ref{pot1}-\ref{pot3}.
In Fig.\ref{pot3} we show the results for $\mu$ and $\mu^{\pm}$ for the symmetric case $J=11/2$
and $N=12$. We see that again the same remarks as for the asymmetric cases hold true. However, we notice
the particular situation that for $\mu$ the exact, BCS, and SCQRPA solutions coincide exactly. This has to do
with the specific symmetries in the half-filled case.   

\begin{figure}                       
\resizebox{0.65 \columnwidth}{!}{\includegraphics{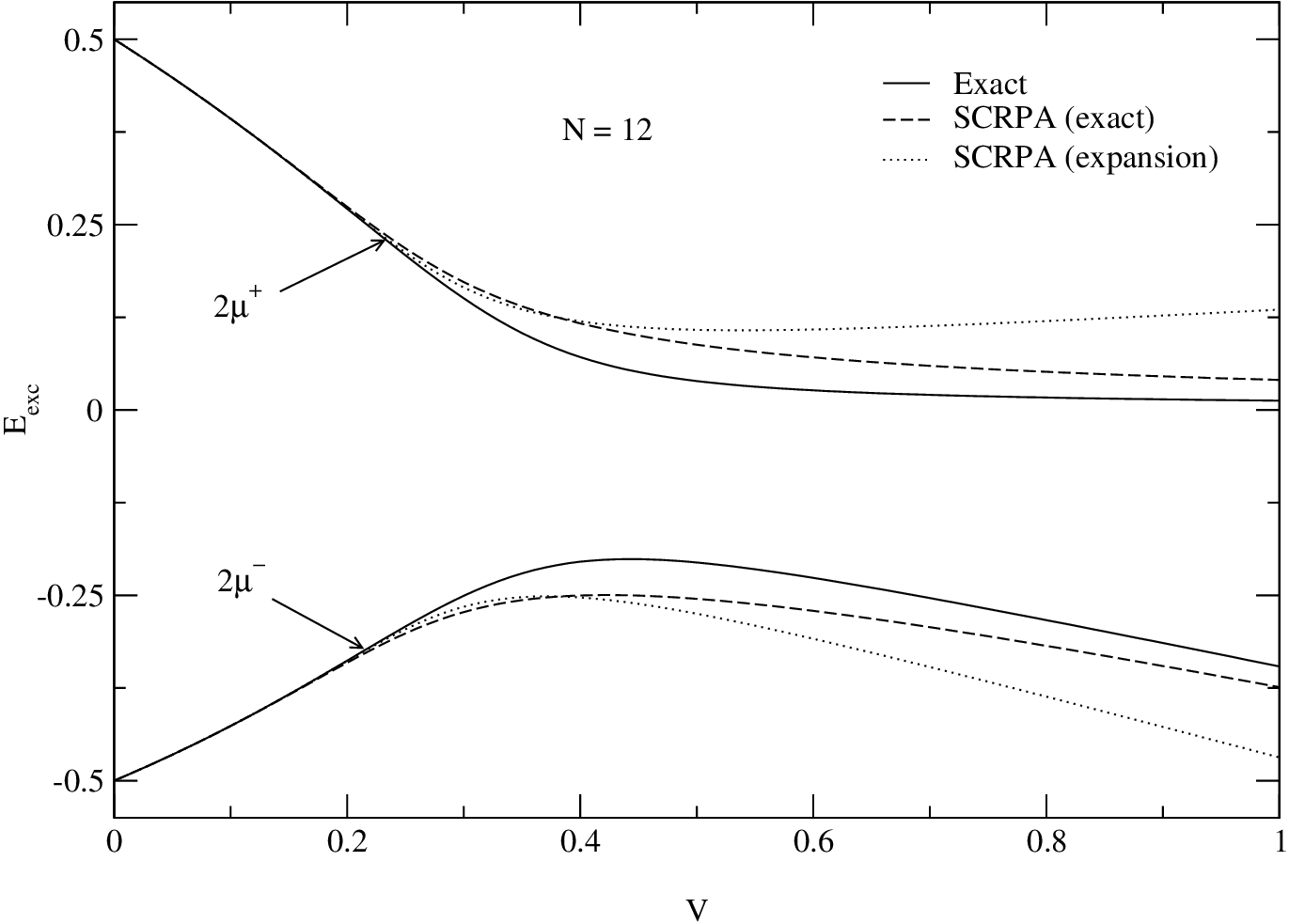}}
\caption{Excitation energies $2\mu^{+}$ (upper lines) and $2\mu^{-}$ (energies are divided by $2\epsilon$)
as a function of the variable $V=g\Omega/2\epsilon$ described in the text and for $N=12$. 
The spin of the levels is $J=11/2$. The full lines correspond to the exact results, the broken 
lines to SCRPA with occupation numbers calculated with the wave function~(\ref{wf}), and dotted 
lines to SCRPA with occupation numbers from the expansion~(\ref{a44}).}
\label{bosonic}
\end{figure}

It is also interesting to show the chemical potentials $2\mu^{+}$ and $2\mu^{-}$ in a symmetric way as done in
\cite{b27}. This also gives us the occasion to study the accuracy of our approximation~(\ref{a44}) and~(\ref{a46}) 
for the occupation numbers. Lets us first of all say that we have here a quite unusual situation for SCRPA : 
the solution in the spherical, i.e. non superfluid basis, exists far into the superfluid regime. 
Usually in other models the solution of SCRPA in the spherical basis can be found up to interaction values 
slightly beyond the mean field transition point but here very reasonable values for the chemical potentials 
$2\mu^{\pm}$ are obtained for all values of $V$ as seen in Fig.\ref{bosonic}. This was also found in the work 
by Passos and al.~\cite{b27}. It should be mentioned, however, that maintaining the spherical basis gives much 
less good results for the ground-state energy as seen in Fig.\ref{full1}. Indeed after the transition point 
the ground-state energy values deviate quite strongly from the exact results. In Fig.\ref{bosonic} we calculate
the expectation values $\langle \hat{N}_{i}\rangle$ and $\langle \hat{N}_{i}\hat{N}_{j}\rangle $ with the exact 
RPA ground-state~\cite{b13}
\beq
|RPA\rangle =\sum_{l=0}^{\Omega}\big(\frac{Y}{X}\big)^{l}\big(A^{\dagger}_{1}\big)^{l}\big(A^{\dagger}_{-1}
\big)^{\Omega -l}|-\rangle,
\label{wf}
\eeq
where $X$, $Y$ are the RPA amplitudes, defined with the addition (P) and removal (R) phonons of the
particle-particle RPA and satisfying the normalization condition $X^{2}-Y^{2}=1$.  
This gives the broken lines. If we calculate the same values from our limited expansion~(\ref{a44}) then the 
dotted lines are obtained. We see that beyond the transition point the solution becomes extremely sensitive 
to approximations. Indeed our approximated values deviate quite a bit from the ones calculated with the full 
wave function $|RPA\rangle$. 
 
\begin{figure}                       
\resizebox{0.65 \columnwidth}{!}{\includegraphics{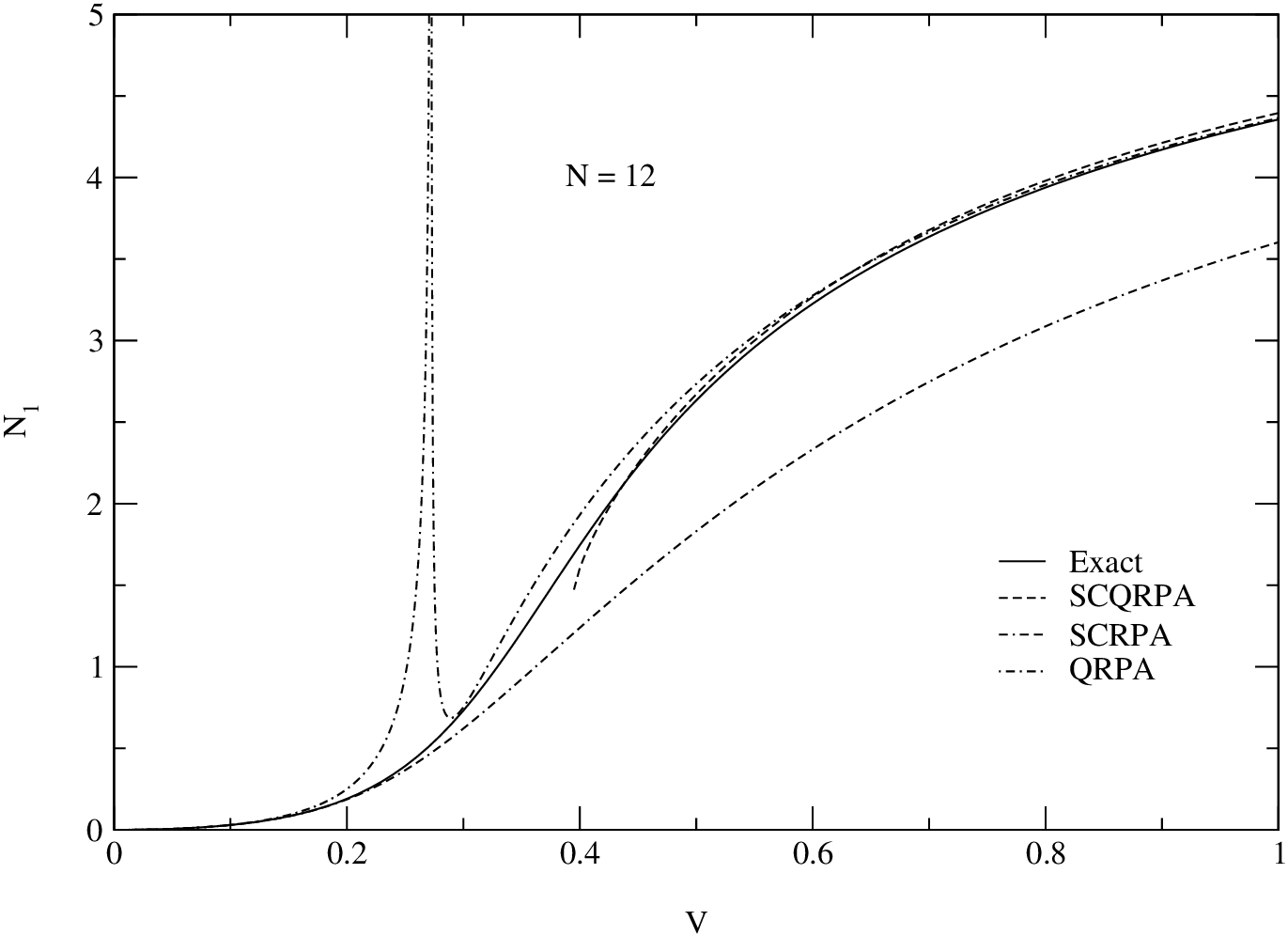}}
\caption{Particle number in the upper level $N_{1}$ as function of the variable $V=g\Omega/2\epsilon$ described
in the text and for $N=12$. The spin of the levels is $J=11/2$.The results refer to exact calculations (solid
line), SCQRPA (dashed line), SCRPA (double-dash dotted line) and QRPA (dot-dashed line).}
\label{upper}
\end{figure}

A quantity which is particularly sensitive to the correct treatment of correlations are the
occupation numbers. For example, for the particle number in the upper level we obtain
\beq
\langle \hat{N}_{1} \rangle=(u_{1}^2-v_{1}^{2})\langle\hat{N}_{q,1}\rangle+2\Omega v_{1}^{2}
\label{a52}
\eeq
and the result is shown in Fig.\ref{upper} for the superfluid and non superfluid regimes. Once again we
see that the change around the phase transition is not continuous. Still with SCQRPA one notices a tremendous
improvement over standard QRPA for which the amplitudes diverge at the critical point. Indeed it
is just in such quantities as occupation numbers where the full superiority of SCQRPA over its linearized
version of QRPA is fully born out. Before finishing this section, we will explain how we proceeded to make the
QRPA and RPA calculation of $\langle\hat{N}_{q,1}\rangle$ in both regions normal and superfluid. We use the first 
order of the bosonic expansion of the $\hat{N}_{q,1}$, i.e. the first order of the expansion shown in~(\ref{a44}), 
where it is sufficient to put $P^{\dagger}_{1}=B^{\dagger}_{1}$. Thus, with the commutation rules~(\ref{a39}), 
we find    
\beq
\langle \hat{N}_{q,1}\rangle=\frac{2(Y^{2}_{1,1}+Y^{2}_{1,2})}{(1+\frac{2}{\Omega}(Y^{2}_{1,1}+Y^{2}_{1,2}))}.
\eeq
In linearizing this expression, we obtain 
\beq
\langle \hat{N}_{q,1}\rangle=2(Y^{2}_{1,1}+Y^{2}_{1,2}). 
\label{num1}
\eeq
It is interesting to detail this calculation, since it is useful to see analytically the QRPA and RPA 
results for the particle number in the upper level close the transition point. It is well known that the two 
excitation modes in the RPA method  converge to zero at the transition point, then the corresponding RPA 
amplitudes tend to infinity, what explains the divergence of $\langle \hat{N}_{1}\rangle$. 
In the superfluid zone, we mention that we neglected the RPA amplitudes corresponding to the Goldstone (spurious) 
mode when we make the calculation of $\langle \hat{N}_{1} \rangle$.        

\begin{figure}                       
\resizebox{0.65 \columnwidth}{!}{\includegraphics{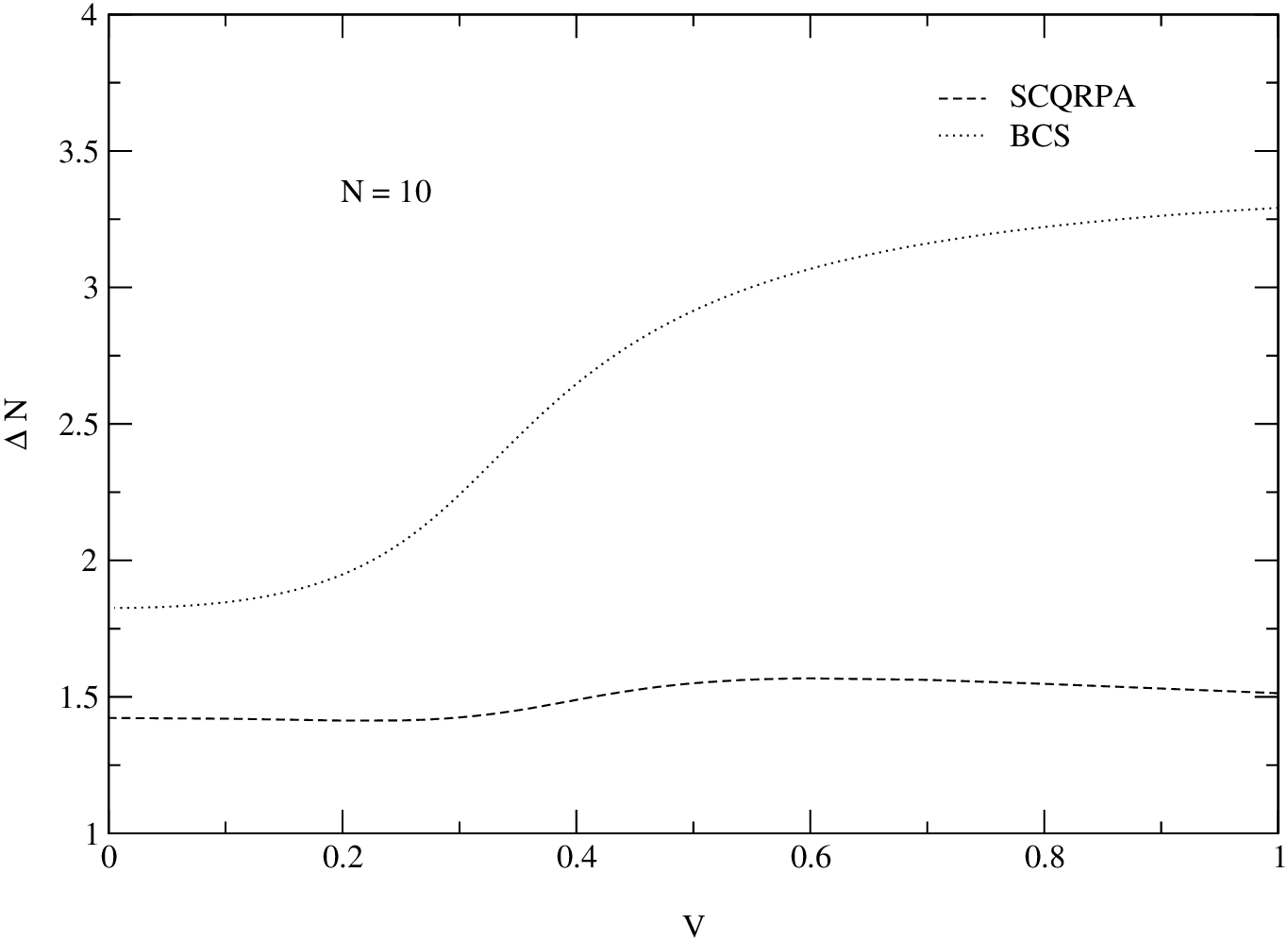}}
\caption{Variance as a function of the variable $V=g\Omega/2\epsilon$ described in the text 
and for particle number $N=10$. The spin of the levels is $J=11/2$. 
The results refer to SCQRPA calculations (dashed line) and BCS (dotted line).}
\label{fluct10}
\end{figure}
\begin{figure}                       
\resizebox{0.65 \columnwidth}{!}{\includegraphics{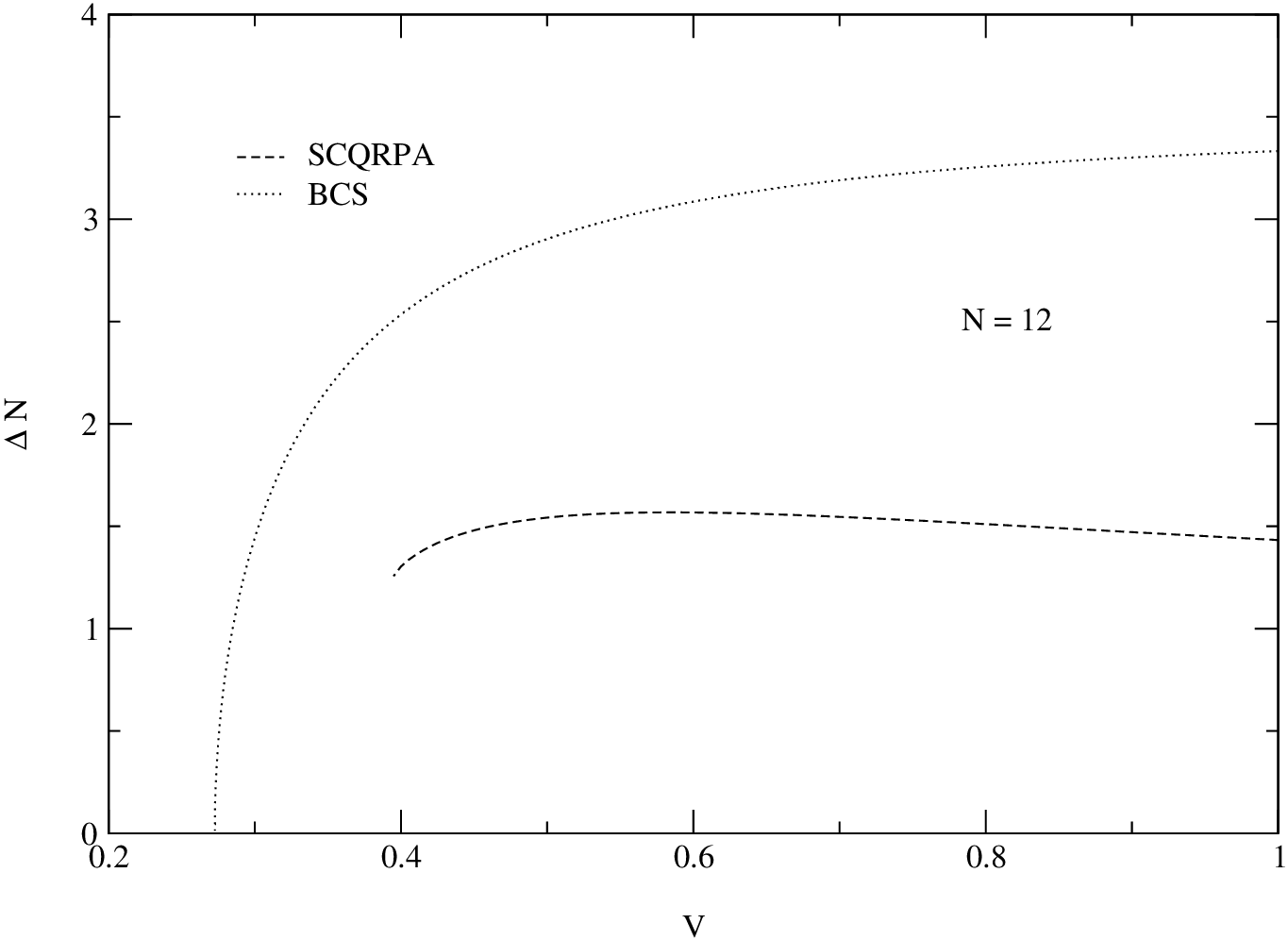}}
\caption{As in Fig.\ref{fluct10} but $N=12$.}
\label{fluct12}
\end{figure}
\begin{figure}                       
\resizebox{0.65 \columnwidth}{!}{\includegraphics{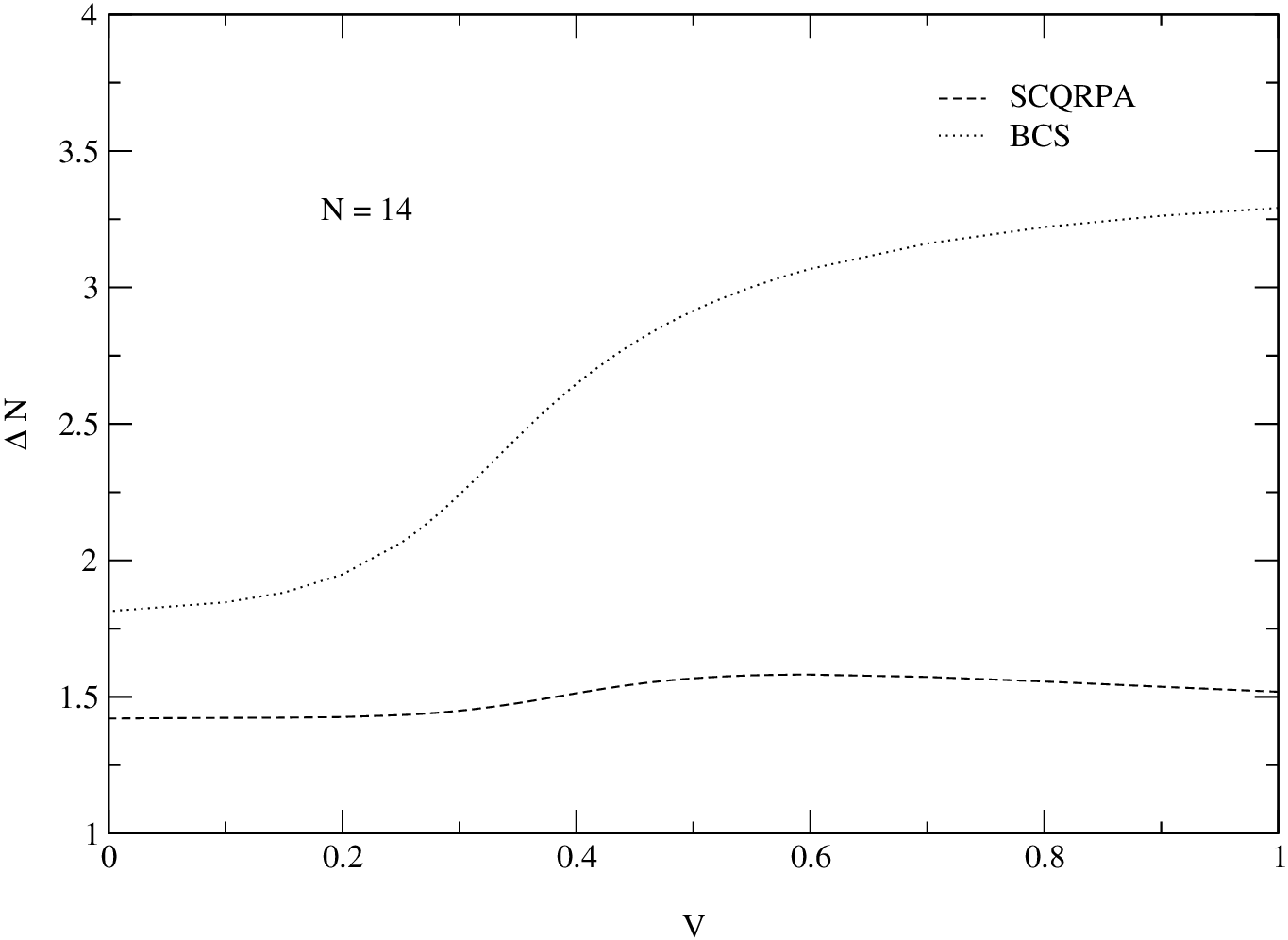}}
\caption{As in Fig.\ref{fluct10} but $N=14$.}
\label{fluct14}
\end{figure}

A constant concern for superfluidity or superconductivity in finite systems is that the quasi-particle
transformation~(\ref{a9}) does not preserve good particle number. Even though one fixes particle
number in the mean with the help of a Lagrange multiplier, the contamination of expectation values
with components which have wrong particle number can be quite important. This is for instance the
case for atomic nuclei. That is why, very early, one has thought of how to improve BCS theory with
respect to particle number conservation. One quite popular approach is to project the BCS wave
function on good particle number. An approximation to this relatively heavy scheme is the
approximate particle number projection by Lipkin-Nogami~\cite{b28}. It is therefore interesting to
investigate how much SCQRPA improves on the spread in particle number. We therefore will calculate
\beq
(\Delta N)^{2}=\langle \hat{N}^{2}\rangle -{\langle \hat{N}\rangle}^{2}
\label{a53}
\eeq
with $\hat{N}=\hat{N}_{1}+\hat{N}_{-1}$ the particle number operator and $\hat{N}_{j}$ given 
by~(\ref{a18}) within SCQRPA. The terms involving bilinear forms in $P^{\dagger}_{j}$, $P_{j}$ are
as usual directly expressed by the RPA amplitudes and for the quasi-particle occupation number operators
we use~(\ref{a44})-(\ref{a46}). Then $\Delta N$ can be calculated and the results for various
configurations are shown in Fig.\ref{fluct10}-\ref{fluct14}. We see that the spread in particle number is
strongly reduced over BCS values reaching typical factors two to three. We, however, see that $\Delta N$ 
even in SCQRPA acquires non vanishing sizable values. This is an expression that particle number is not 
completely restored. We will see in section \ref{senio} how one eventually can improve on this. 
We also tried to evaluate $\Delta N$ in standard QRPA in applying a lowest order bosonization of
the expression. However, due to the non-normalizable Goldstone mode we ran into troubles with this 
procedure and could not reach a definite conclusion on this point.

\begin{figure}                       
\resizebox{0.65 \columnwidth}{!}{\includegraphics{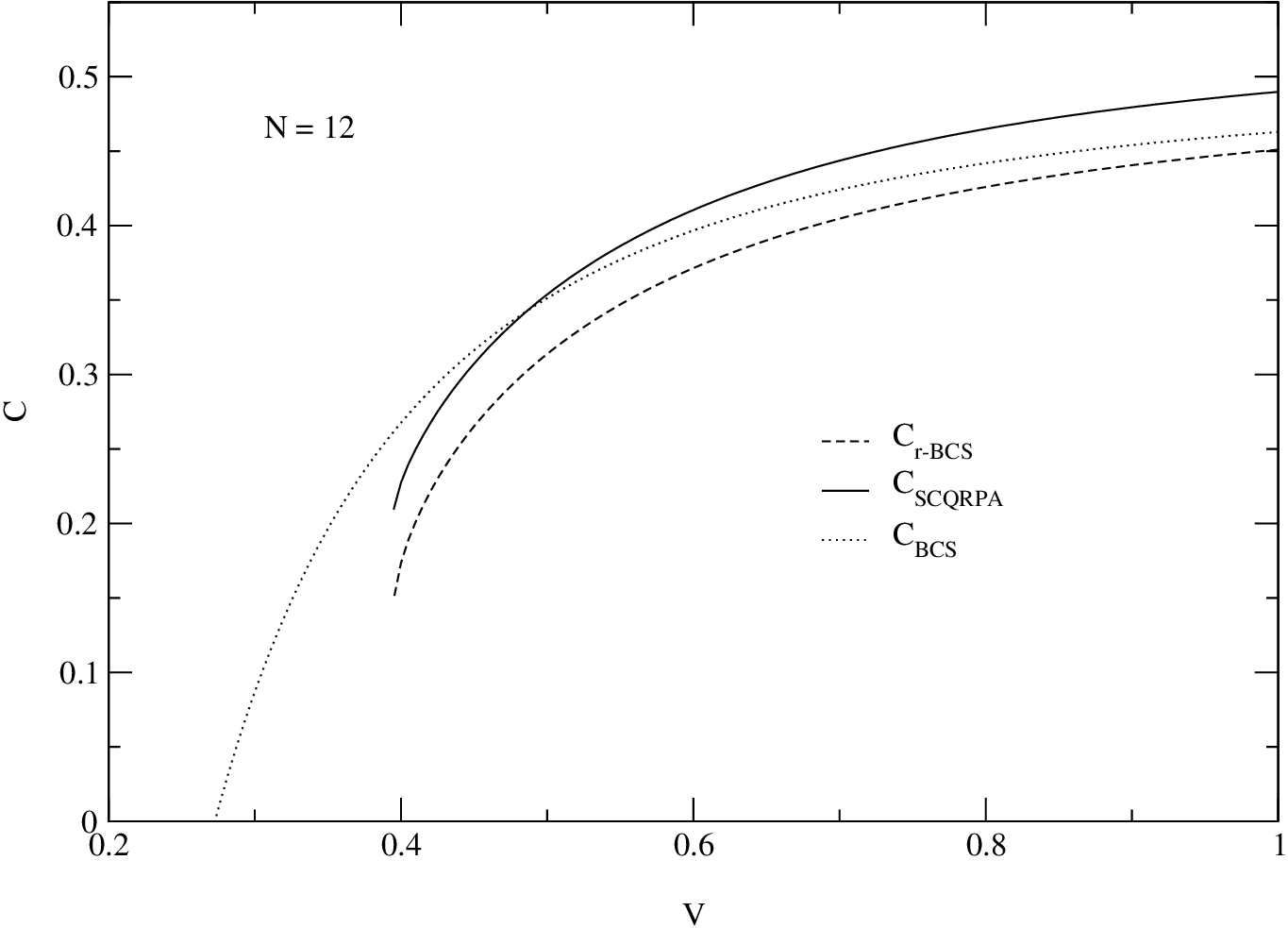}}
\caption{Correlation function ${\cal C}$ as a function of the variable $V=g\Omega/2\epsilon$ described 
in the text and for particle number $N=12$. The spin of the levels is $J=11/2$.The results refer to 
SCQRPA (solid line), renormalized BCS (dashed line) and standard BCS (dotted line).}
\label{delta12}
\end{figure}

\begin{figure}                       
\resizebox{0.65 \columnwidth}{!}{\includegraphics{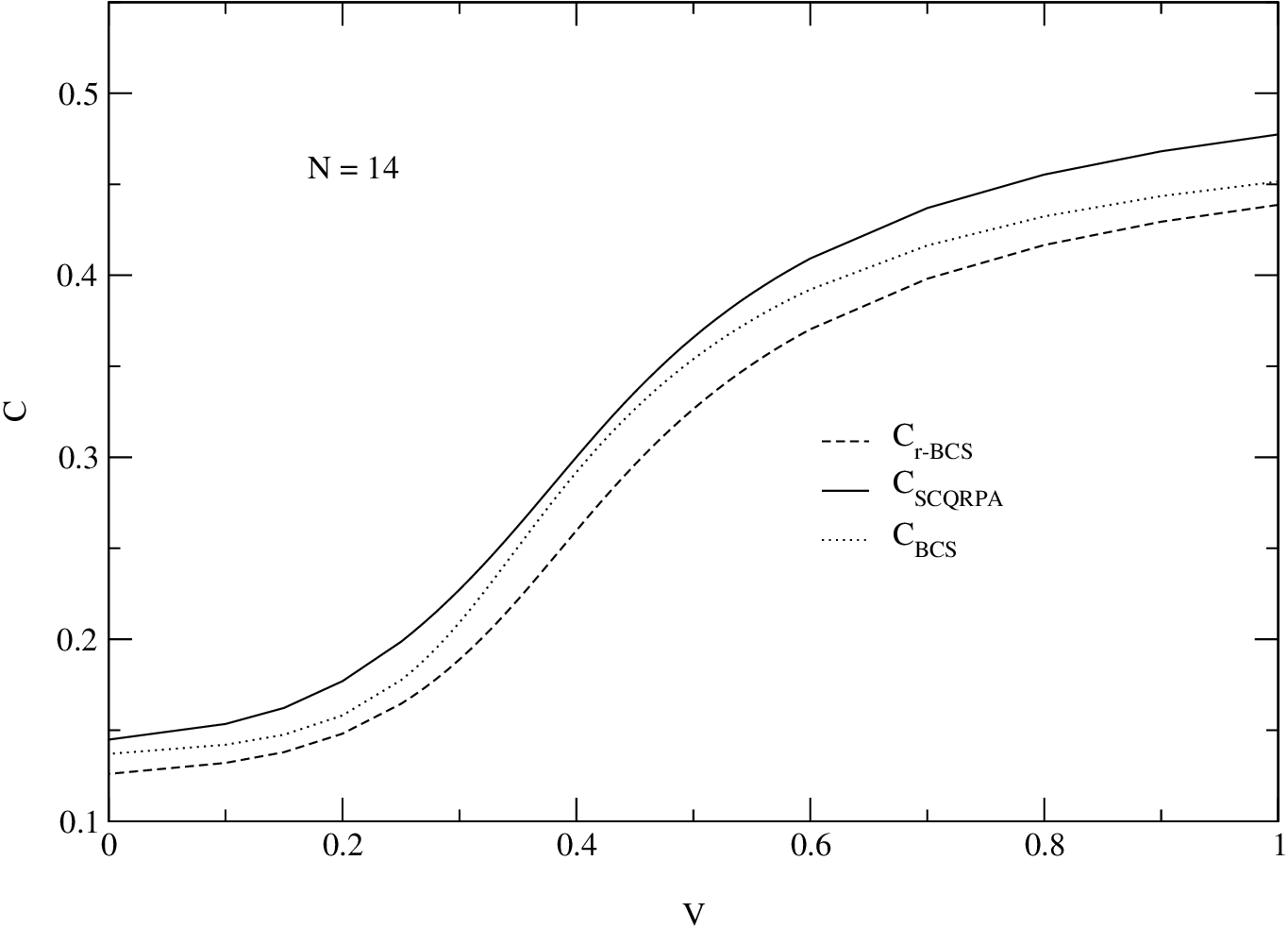}}
\caption{As in Fig.\ref{delta12} but $N=14$.}
\label{delta14}
\end{figure}

Another interesting aspect which can be studied with our model is the question whether the pairing
correlations, with respect to BCS, have been enhanced or weakened due to the SCQRPA correlations.
To this end we define the following quantal expression for the correlation function
\beq
{\cal C} =\frac{1}{\Omega}\sum_{j=\pm 1}\bigg( \langle A^{\dagger}_{j}A_{j}\rangle
-\frac{1}{4\Omega^{2}}\langle\hat{N}_{j}\rangle
\langle\hat{N}_{j}\rangle\bigg).  
\label{corr}
\eeq
This expression reduces to the following expression when evaluated with the BCS ground-state
\beq
{\cal C}_{BCS}=\sum_{j=\pm 1} u^{2}_{j}v^{2}_{j}.
\label{uv}
\eeq
Multiplying~(\ref{uv}) with $g^{2}\Omega^{2}$ yields the standard BCS gap squared. We, however, refrain from
multiplying~(\ref{corr}) or~(\ref{uv}) with $g^{2}\Omega^{2}$, since the renormalized gap from~(\ref{a34}) is
level dependent. Often also~(\ref{corr})  is given in a non diagonal form~\cite{b34} but having difficulties 
to express non diagonal densities with SCQRPA amplitudes we will not consider the non diagonal form here. We 
therefore evaluate~(\ref{corr}) in three approximations : we can express~(\ref{corr}) in terms of 
$P^{\dagger}_{j}$, $P_{j}$ and $\hat{N}_{q,j}$ operators and then take the expectation value with the SCQRPA 
ground-state. The equations~(\ref{a23}), (\ref{a44}) and~(\ref{a45}) then allow to express ${\cal C}$ in terms
of the SCQRPA amplitudes $X_{j,\nu}$, $Y_{j,\nu}$. We will call this ${\cal C}_{SCQRPA}$. We also 
evaluate~(\ref{uv})  in the standard BCS approximation which is~(\ref{corr}). However, we also calculate~(\ref{uv})
with $u_{j}$, $v_{j}$ amplitudes from the renormalized BCS (r-BCS) theory, i.e. from~(\ref{a33}) with $\Delta_{i}$ 
solution of~(\ref{a34}). The results are shown in Fig.\ref{delta12} and Fig.\ref{delta14} for $N=12$
and $N=14$ respectively (the case $N=10$ gives exactly same results as $N=14$). We see that r-BCS gives 
with respects to BCS less correlations. Eventually this suppression of pairing can be put into analogy with
gap suppression in infinite neutron matter from renormalized theories~\cite{b3} (see discussion in the 
introduction). However, the suppression of pairing correlation in r-BCS is misleading in our model, since, on
the contrary, the full SCQRPA gives mostly an enhancement of pair correlations with respect to BCS. It is not 
obvious whether this conclusion can be taken over to the infinite matter case. It may, however, be indicated that 
the renormalized gap equations from screening (RPA) type correlations should be carefully treated consistently 
with the evaluation of two body correlation function before definite conclusions can be reached.              

\begin{table*}
\caption{
\label{table}
Results for the ground-state energy (in arbitrary units) vs the variable $V=g\Omega/2\epsilon$ described 
in the text. The spin of the levels is $J=19/2$ and the number of particles is $N = 20$.}  
\begin{ruledtabular}
\begin{tabular}{ c c c c c c }
V & Exact & SCQRPA & BF-RPA & QRPA & BCS \\
\hline
-0.50 & -18.55446 & -18.55410 & -18.55360 & -18.56890 & -18.00000 \\
-0.45 & -18.66849 & -18.66821 & -18.66784 & -18.67924 & -18.20000 \\
-0.40 & -18.78706 & -18.78686 & -18.78660 & -18.79474 & -18.40000 \\
-0.35 & -18.91072 & -18.91058 & -18.91040 & -18.91592 & -18.60000 \\
-0.30 & -19.04010 & -19.04001 & -19.03990 & -19.04338 & -18.80000 \\
-0.25 & -19.17600 & -19.17594 & -19.17588 & -19.17787 & -19.00000 \\
-0.20 & -19.31939 & -19.31936 & -19.31933 & -19.32031 & -19.20000 \\ 
-0.15 & -19.47153 & -19.47151 & -19.47150 & -19.47188 & -19.40000 \\ 
-0.10 & -19.63406 & -19.63405 & -19.63405 & -19.63415 & -19.60000 \\
-0.05 & -19.80919 & -19.80919 & -19.80919 & -19.80919 & -19.80000 \\
 0.00 & -20.00000 & -20.00000 & -20.00000 & -20.00000 & -20.00000 \\ 
 0.05 & -20.21101 & -20.21101 & -20.21101 & -20.21102 & -20.20000 \\
 0.10 & -20.44921 & -20.44918 & -20.44917 & -20.44953 & -20.40000 \\
 0.15 & -20.72625 & -20.72599 & -20.72593 & -20.72899 & -20.60000 \\
 0.20 & -21.06339 & -21.06130 & -21.06100 & -21.08080 & -20.80000 \\
 0.25 & -21.50260 & -21.48733 & -21.48640 & -21.64174 & -21.00000 \\
 0.30 & -22.12491 & -22.03638 & -22.03620 & -22.13484 & -21.37193 \\
 0.35 & -23.03321 & -22.80453 & -22.77899 & -22.99299 & -22.21880 \\
 0.40 & -24.24609 & -23.99588 & -23.97001 & -24.21285 & -23.37895 \\
 0.45 & -25.68929 & -25.42829 & -25.39821 & -25.65779 & -24.74795 \\
 0.50 & -27.29077 & -27.01885 & -26.98390 & -27.25633 & -26.26316 \\
\end{tabular}
\end{ruledtabular}
\end{table*}

\section{Comparison with other recent works}
\label{compar}
The two level pairing model has recently served as a testing ground for various generalizations of
BCS theory. In spirit the work which comes closest to the present is the one of Sambataro and Dinh 
Dang~\cite{b23}.
Instead of treating quasi-particle pair operators directly as we do here, they bosonize them (with a
method developed in~\cite{b23}) and expand the Hamiltonian~(\ref{a1}) in terms of these Bosons up to fourth
order. A Bogoliubov transformation of the Boson operators quite analogous to our Bogoliubov
transformation of Fermion pair operators~(\ref{a9}) is then performed and the corresponding non linear
Hartree-Fock-Bogoliubov equation are written down. Again they are quite analogous to our SCQRPA
equations. The coefficients of quasi-particle transformation are obtained, as usual, in minimizing
the ground-state energy (see also our procedure~(\ref{a29})) with respect to the transformation
coefficients. As in our case equations are obtained which couple back to the bosonic HFB, i.e. the RPA
amplitudes. The coupled system of equations for fermionic and bosonic transformation amplitudes is
then solved self-consistently. For better comparison we actually, on purpose, have chosen most of the
configurations in $J$ and $N$ the same as in~\cite{b23}. Since in~\cite{b23} $J=11/2$, what is a rather high
degeneracy of the levels, the Fermion pair operators are quite collective and a bosonization
certainly is a valid procedure. Not unexpectedly, therefore, the results of the
present work are very close to the ones presented in~\cite{b23}. A detailed comparison shows that our
results are systematically closer to the exact ones by a very small amount. 
This may be due to the fact that we never bosonize and treat the Fermion pair commutation rules exactly 
but the difference is too small for drawing any definite conclusion. In~\cite{b23}, Sambataro
and al. show an explicit comparison of results referring to the symmetric case with $J=19/2$. We also can 
make such a comparison for the ground-state energy referring to the same configuration. It is given 
in TABLE.\ref{table}, where we show the results for four different many-body approaches : 
our approach (SCQRPA), approach of Sambataro and al. (BF-RPA)~\cite{b23}, standard Quasi-Particle RPA 
(QRPA)~\cite{b24} and the BCS 
method~\cite{b18}. Of course, we recall that in our approach, we use the Self-Consistent particle-particle RPA 
in the normal fluid zone, while, in the superfluid region we use the generalized version of SCRPA that is SCQRPA. 
In order to accentuate the differences one would have to go to configurations with much lower degeneracies 
where the constraints from the
Pauli principle become much more severe. For example the SCRPA approach has been applied to the case
$J=1/2$ with $N=2$ in~\cite{b7}  and the exact result was recovered. It would be interesting to see how the
approach with the bosonization~\cite{b23} performs in that case. In spite of being very similar in general
spirit to the work in~\cite{b23} we have solved and considered several additional problems which remained open
in~\cite{b23}.
In first place this concerns the low lying eigenvalue of SCQRPA. No interpretation of this important
root was given in~\cite{b23}. We, however, suppose that the results in~\cite{b23} for this state (no numerical
values have been given) can be equally interpreted as the difference $2(\mu^{+}-\mu^{-})$ as in
our case. Another quantity which was not considered in~\cite{b23} is the number fluctuation. Again we
believe that corresponding values would be close to the ones found here. Also the transition to the
non superfluid regime has not been treated in~\cite{b23}. However, probably all these aspects will be 
quite similar in both approaches as long as a bosonization of the Fermion pair operators is valid. 
We think, however, that it does not cost much to avoid bosonization altogether as with the SCQRPA approach. 

Also in the work by Passos and al.~\cite{b27} the SCRPA method was applied to the present model. However,
only the non superfluid formulation, i.e. SCRPA, was studied. The results are quite analogous to ours. 
In addition in~\cite{b27} a further approximation, half way between RPA and SCRPA, the so-called renormalized 
RPA (r-RPA) where only the single-particle occupation numbers are allowed to be affected by ground-state
correlation, has been considered. The astonishing finding there was that the exact occupation numbers
are almost perfectly reproduced over the whole range of the coupling constant with r-RPA but not with
SCRPA which under shoots the correlations. This was interpreted in~\cite{b27} as a positive feature of r-RPA
over SCRPA. We can not follow this conclusion from our experience with SCRPA in this and other works~\cite{b7}. 
As we outlined above, any relax of the severe constraints of the Pauli principle respected in SCRPA
will inevitably lead to more correlations as there should be. It can happen by accident
that if one relaxes the Pauli principle just by the right amount that one falls more or less on the exact
values. This is what happened for the QRPA ground-state energy discussed above and apparently it is also
what happens for the occupation numbers in r-RPA. However, we think this result can not be generalized and
for other models or physical situations the scenario may be completely different. The only really trust
worthy theory is the full SCRPA approach, since it can be derived from a variational principle. If the
results are not good, one must improve on SCRPA (i.e. include e.g. higher configurations) and not
approximate it.

In the work by Hagino and Bertsch~\cite{b24} the QRPA approach is advocated. This in the spirit to have a
numerically viable alternative to projected BCS and the method by Nogami-Lipkin~\cite{b28}. It is certainly
true that in realistic cases SCRPA is numerically very demanding, though probably not impossible to solve
with modern computers. Then, of course, in a first step it is worthwhile to investigate
standard QRPA. This is for instance true if one intends to do large scale calculations for a great
number of nuclei~\cite{b24}. However, one should always remember that standard QRPA may have quite important
failures which certainly will be most prominent in situations where the system is close to a phase
transition.   

\section{The question of a second constraint on the particle number variance}
\label{senio}
As we have seen above, with respect to BCS the SCQRPA reduces the spread in particle number by an
important factor. However, the variance $\Delta N$ is still appreciable and one can ask the question
whether it is not possible to further improve the theory on this point. A natural idea which comes to
mind is that instead of fixing only $\langle \hat{N} \rangle =N$, one could at the same time fix $\langle
\hat{N} \hat{N}\rangle =N^{2}$ with a second Lagrange multiplier. Since in SCQRPA the number of
variational parameters is largely increased with respect to BCS one could imagine that there is
indeed enough freedom for constraining the particle number fluctuation to zero. The Hamiltonian to be
considered is therefore
\beq
H^{\prime}=H-\mu_{1}\hat{N}-\mu_{2}{\hat{N}}^{2}.
\label{a54}
\eeq
Let us immediately give our conclusion : in the two level pairing case we could not find a solution to
this problem.
The system of non linear equations with the two constraints $\mu_{1}$ and $\mu_{2}$ is quite complex
and in spite of considerable numerical effort we did not have success to get the solution converged.
We were not able to decide whether the difficulty is purely numerical or whether there is a principal
problem. In fact we were at first encouraged by results we obtained in the one level pairing case 
(the seniority model). The outcome of employing the second constraint was that the one level model was
solved exactly. In spite of being a somewhat trivial model which certainly limits the conclusions, it may
be interesting to show how this goes.
The Hamiltonian to be considered is now
\beq
H^{\prime}=H-\mu_{1}\hat{N}-\mu_{2}{\hat{N}}^{2}.
\label{a55}
\eeq
where $\mu_{1}$ and $\mu_{2}$ are two Lagrange multipliers fixing $\langle \hat{N} \rangle = N $ and 
$ \langle \hat{N}^{2}\rangle = N^{2}$ and 
\begin{equation}
H=- g\Omega A^{\dagger}A
\label{a56}
\end{equation}
with in analogy to~(\ref{a2})
$A^{\dagger}=\frac{1}{\sqrt{\Omega}}\sum_{m>0}a^{\dagger}_{m}a^{\dagger}_{-m}$, and where we put the
origin of energy at the single-particle level. As in the two level case we transform to quasi-particles
and with only one level the SCQRPA equation reduce to a $(2 \times 2)$ eigenvalue problem
\begin{equation}
\pmatrix{ A & B \cr -B & -A \cr }
\pmatrix{ X \cr  Y \cr } = E \pmatrix{ X \cr Y \cr };\quad X^{2}-Y^{2}=1
\label{a57}
\end{equation}
where in analogy with~(\ref{a27})
\bwt\bea
A&=&2(g-4\mu_{2})\bigg\{(2v^{2}(1-v^{2})-1)XY-v^{2}(1-v^{2})(1+2Y^{2}-\Omega\frac{1-2\frac{\langle
\hat{N}_{q}\rangle}{\Omega}+\frac{\langle\hat{N}_{q}^{2}\rangle}{\Omega^{2}}}{1-\frac{\langle
\hat{N}_{q} \rangle}{\Omega}})\bigg \}, \label{a58}\\
B&=&2(g-4\mu_{2})\bigg\{(6v^{2}(v^{2}-1)+1)XY-v^{2}(1-v^{2})(1+2Y^{2}-\Omega\frac{1-2\frac{\langle
\hat{N}_{q}\rangle}{\Omega}+\frac{\langle\hat{N}_{q}^{2}\rangle}{\Omega^{2}}}{1-\frac{\langle
\hat{N}_{q} \rangle}{\Omega}})\bigg\}.
\label{a59}
\eea\ewt
The Bogoliubov transformation to quasi-particles is obtained as for the case of two levels
\begin{equation}
\pmatrix{ h & \Delta \cr \Delta & -h \cr }
\pmatrix{ u \cr  v \cr } = \tilde{\epsilon} \pmatrix{ u \cr v \cr }
\label{a60}
\end{equation}
with in close analogy to the expression~(\ref{a29}) of the two levels
\bwt\beas
h&=&-(g+4\mu_{2}\Omega)v^{2}-\mu_{1},  \slabel{h}\\
\Delta&=&(g\Omega+4\mu_{2})uv-(g-4\mu_{2})uv \bigg\{ 2(XY+Y^{2})+\frac{\langle
\hat{N}_{q}\rangle
-\frac{\langle \hat{N}_{q}^{2}\rangle}{\Omega}}{1-\frac{\langle
\hat{N}_{q}\rangle}{\Omega}}\bigg\},  \slabel{Delta}\\
\tilde{\epsilon}&=&\sqrt{h^{2}+\Delta^{2}} \slabel{epsilon}.
\label{a61}
\eeas\ewt
In addition to the SCQRPA equations~(\ref{a57}) we have two further equations which, in principle,
allow us to find the Lagrange multipliers $\mu_{1}$ and $\mu_{2}$ (see, however, below)
\bwt\begin{subeqnarray}
N&=&\langle \hat{N}\rangle =(u^{2}-v^{2})\langle\hat{N}_{q}\rangle+2\Omega v^{2},\slabel{seq1}\\
N^{2}&=&\langle \hat{N}^{2}\rangle=(u^{2}-v^{2})^{2}\langle\hat{N}_{q}^{2}\rangle 
+8\Omega u^{2}v^{2}(1-\frac{\langle\hat{N}_{q}\rangle} 
{\Omega})(XY+Y^{2})+4\Omega v^{2}(u^{2}+\Omega v^{2}) \cr
&+&4v^{2}(\Omega(u^{2}-v^{2})-u^{2})\langle\hat{N}_{q}\rangle \slabel{seq2}
\label{a62}
\end{subeqnarray}\ewt
We again see that eqs~(\ref{a62}) reduce to the standard expressions, once, as in
the HFB approximation, we pose $Y=\langle\hat{N}_{q}\rangle=\langle\hat{N}_{q}^{2}\rangle=0$. In the
case of the seniority model the number equation~(\ref{seq1}) in the HFB approximation
determines the amplitudes $u, v$ and then no freedom is left to impose 
$\Delta N=\sqrt{\langle \hat{N}^{2} \rangle -{\langle \hat{N} \rangle}^{2}}=0$. However, 
in the more general approach of SCQRPA there is more freedom and, as we will see, we will be able to 
satisfy the relation $\Delta N= 0$ as well. For $\hat{N}_{q}$ and $\hat{N}^{2}_{q}$ we have the same relation
as in~(\ref{a44}) and~(\ref{a45}). Again the system of equations is therefore closed.

Usually the number equation~(\ref{seq1}) and~(\ref{seq2})
are to be used for the determination of the chemical potential $\mu_{1}$ and the
second Lagrange multiplier $\mu_{2}$ and the mean field equations for the amplitudes
$u, v$ and $X, Y$. In the present case it is, however, more convenient to invert the role of mean
field and number equations, since eqs~(\ref{a61}) do not depend on the Lagrange multipliers and
therefore readily allow to determine $v^{2}$ and $Y^{2}$ as a function of the particle number $N$.
Inversely the two mean field eqs~(\ref{a57}),~(\ref{a60}) are linear in $\mu_{1}$
and $\mu_{2}$ and for instance it is seen that~(\ref{a57}) directly yields
\begin{equation}
\mu_{2}=\frac{g}{4}  
\label{a63}
\end{equation}
independent of the particle number $N$. Considering, the well known exact
expression for the ground-state energy of the model~\cite{b18}
\begin{equation}
E_{0}=-\frac{g}{2}(\Omega+1)N+\frac{g}{4}N^{2}
\label{a64}
\end{equation}
we see from $\mu_{2}=\frac{\partial^{2} E_{0}}{\partial N^{2}}$ that~(\ref{a63}) gives
the exact value for the second Lagrange multiplier $\mu_{2}$. For the chemical potential $\mu_{1}$ we
obtain from~(\ref{a60})
\bwt\begin{equation}
\mu_{1}=(g-4\mu_{2})\bigg\{(\Omega-1)v^{2}+(1-2v^{2})(XY+Y^{2}+\frac{1}{2}\frac{\langle\hat{N}_{q}
\rangle-\frac{\langle\hat{N}_{q}^{2}\rangle}{\Omega}}{1-\frac{\langle
\hat{N}_{q}\rangle}{\Omega}})\bigg\}-\frac{g}{2}\Omega-2\mu_{2}.
\label{a65}
\end{equation}\ewt
With relation~(\ref{a60}) this gives $\mu_{1}=-\frac{g}{2}(\Omega+1)$ which again
is the exact value.
Furthermore, with~(\ref{a63}) we have from~(\ref{a58}),~(\ref{a59}) that $ A=B=0
$ and therefore the RPA eigenvalue $E=0$. This means that, as in standard QRPA, SCQRPA yields 
a Goldstone mode at zero energy.This feature is very rewarding, since it signifies that the particle 
number symmetry is exactly restored. \\
It is well known that restoration of good particle number implies in this very simple model case that
the model is solved exactly~\cite{b18}. We have already seen that one obtains the exact values for 
$\mu_{1}$ and $\mu_{2}$. We now will show that one also obtains the exact value for the ground-state
energy (and therefore for the whole band of ground-state energies). This goes as follows.
For the expectation value of $H$ of eq~(\ref{a56}) in the RPA ground-state, using the
analogous relations~(\ref{a20}) and~(\ref{a23}) for this case and the quasi-particle
representation for $H$, we can write 
\bwt\bea
E_{0}&=&\langle H \rangle \cr
&=&-\frac{g}{2}(\Omega+1)\{(1-2v^{2})\langle \hat{N}_{q} \rangle +2\Omega v^{2}\}
+\frac{g}{4}\bigg\{(1-2v^{2})^{2}\langle \hat{N}_{q}^{2} \rangle \cr
&+&[4v^{2}(\Omega(1-2v^{2})-1+v^{2})-8(1-v^{2})v^{2}(XY+Y^{2})]\langle \hat{N}_{q} \rangle \cr
&+&8\Omega(1-v^{2})v^{2}(XY+Y^{2})+4\Omega v^{2}(1-v^{2}+\Omega v^{2})\bigg\}. 
\label{a66}
\eea\ewt
In this expression we have used the Casimir relation for this case $4Y^{2}(\Omega- \langle \hat{N}_{q} \rangle)=
2(\Omega+1) \langle \hat{N}_{q} \rangle-\langle \hat{N}_{q}^{2} \rangle$ which follows from
(\ref{casimir}). Using the expression for $N$ and $ N^{2}$ of~(\ref{seq1}) and~(\ref{seq2}) once 
more, we see that the exact expression~(\ref{a64}) is recovered. It should be mentioned that because of
the simplicity of the model also the Lipkin-Nogami approach~\cite{b28} solves the model exactly.
\section{Conclusion}
\label{concl} 
In this work we extended for the first time the Self-Consistent RPA theory (SCQRPA) to the superfluid
case for a model with more than one level. Indeed in~\cite{b12} SCQRPA was already applied to the seniority
Model but this only allowed to study rotation in gauge space whereas intrinsic excitations
("$\beta$-vibrations") are absent in the $0^{+}$-sector of the seniority Model. We have considered the two level
version of the pairing Hamiltonian with arbitrary degeneracies and fillings of the levels. We mostly
considered the case $J=11/2$ for the upper and lower levels. This configuration was chosen in order to
have a better comparison with the work by Sambataro and Dinh Dang~\cite{b23} which in many aspects is quite
analogous to ours. Indeed SCQRPA can be considered as a Bogoliubov transformation among
quasi-particle pair operators $\alpha^{\dagger}\alpha^{\dagger}$ and $\alpha\alpha$ whereas in~\cite{b23} the
quasi-particle pair operators were replaced by ideal bosons $\alpha^{\dagger}\alpha^{\dagger} \backsim
B^{\dagger}$ and then a Bogoliubov transformation among these boson operators was applied while the
pairing Hamiltonian was also bosonized up to fourth order. For such collective pairs as they are formed in
$J=11/2$ shells a bosonization seems indeed valid and as expected our results are very close to
the ones given in~\cite{b23}, even though they are consistently slightly better. This could be due to the
fact that in SCQRPA one never bosonizes and rather all constraints from Pauli principle are fully kept.
However, we do not want to attribute too much importance to these differences which only could become
relevant for cases where a bosonization fails. On the other hand in our work considerably more issues
were studied. In first place this concerns the physical interpretation and identification of the low
lying state in SCQRPA. This state corresponds to the Goldstone mode in standard QRPA. However in SCQRPA
this state comes at finite energy and reproduces very precisely the difference $2(\mu^{+}-\mu^{-})$ of
the chemical potentials of the exact solution. We also evaluated the fluctuation $\Delta N$ of the
particle number and showed that with respect to the fluctuation in BCS theory there is a strong
improvement. However, particle number symmetry is still not entirely restored. In spite of this shortcoming 
for $\Delta N$, for other quantities SCQRPA is always superior to BCS and QRPA approaches as explained in 
the main text. In fact the situation with respect to the
particle number symmetry is somewhat particular and not encountered in other cases of spontaneously
broken symmetries. For example in the case of rotation the angular momentum operator $L_{z}$ has no
contributions which are hermitian in the deformed basis and then the Goldstone mode also comes in the case
of SCRPA~\cite{b32}. In order to improve on the restoration of particle number symmetry we also investigated
the possibility of fixing $\langle \hat{N}^{2}\rangle=N^{2}$ with a second Lagrange multiplier.
Whereas in the one level pairing model we could show analytically that this solves the model exactly,
in the two level model we could not find a numerical solution of the system of equations. It remained
unclear whether this is due to some fundamental problem or just a numerical difficulty.

We also discuss carefully in this work the transition from the non superfluid regime to the superfluid
one. We for instance pointed out that the transition from SCRPA to SCQRPA is not continuous and in
fact in both regimes quite different physical excitations are described. This also can be seen looking
at the ground-state energies as a function of the coupling constant. In the transition region there is
no continuous transition between the SCRPA and SCQRPA values but it is definitively seen that the SCRPA values 
for the ground-state energies deviate quite strongly from the exact values after the phase transition whereas
SCQRPA stays close to them.

In conclusion we can say that we have applied with very promising success for the first time SCQRPA to
a more level pairing situation where, at least for the $0^{+}$-sector, all the complexity of a more realistic
situation is present. It could be interesting to extend this work to the description of ultrasmall superconducting 
metallic grains for which the many-level picket fence model seems appropriate~\cite{b7}.

\begin{acknowledgments}
We would like to thank M. Sambataro for useful informations and clarifications concerning Ref.~\cite{b23}. 
One of us (P.S.) specially acknowledges many useful and elucidating discussions with J. Dukelsky. We 
appreciated discussions and interest in this work by G.R\"opke.
\end{acknowledgments}

\appendix
\section{Some useful Mathematical Relations}
\label{average}
At first, we will explain how we calculated the expectation values of type 
$\langle P^{\dagger}_{j} P_{j^{\prime}}\rangle $; 
we recall that $j=\pm 1$. From~(\ref{a23}), i.e. the inversion of the excitation operators $Q_{\nu}$, we can write,
\bea
\bar{P}_{j}&=&\sum_{\nu=1, 2}X_{j,\nu}Q_{\nu}+Y_{j,\nu}Q^{\dagger}_{\nu}, \cr
\bar{P}^{\dagger}_{j}&=&\sum_{\nu=1, 2}X_{j,\nu}Q^{\dagger}_{\nu}+Y_{j,\nu}Q_{\nu}.
\eea
Let us calculate for example $\langle P^{\dagger}_{j} P_{j^{\prime}} \rangle $,
\bwt\beq
\langle P^{\dagger}_{j} P_{j^{\prime}} \rangle = \sqrt{1-\frac{\langle \hat{N}_{q,j^{\prime}}\rangle}{\Omega}}
\bigg(\sum_{\nu=1, 2}X_{j^{\prime},\nu}\langle P^{\dagger}_{j}
Q_{\nu}\rangle + Y_{j^{\prime},\nu}\langle P^{\dagger}_{j}Q^{\dagger}_{\nu}\rangle \bigg)
\eeq\ewt
the first term in the rhs, is zero since $ Q_{\nu}|\rangle=0 $.
Therefore, we obtain,
\bwt\bea
\langle P^{\dagger}_{j} P_{j^{\prime}} \rangle &=&\sqrt{1-\frac{\langle \hat{N}_{q,j^{\prime}}
\rangle}{\Omega}}\sum_{\nu=1, 2}Y_{j^{\prime},\nu} \langle P^{\dagger}_{j}Q^{\dagger}_{\nu}\rangle, \cr
  &=&\sqrt{1-\frac{\langle \hat{N}_{q,j^{\prime}}\rangle}{\Omega}}\sqrt{1-\frac{\langle
  \hat{N}_{q,j}\rangle}{\Omega}}\sum_{\nu=1, 2}Y_{j^{\prime},\nu}
  \langle [\bar{P}^{\dagger}_{j}, Q^{\dagger}_{\nu}]\rangle
\eea\ewt
using the definition of the excitation operators $Q^{\dagger}_{\nu}$~(\ref{a21}), we find,
\bwt\bea
\langle P^{\dagger}_{j} P_{j^{\prime}} \rangle &=& -\sqrt{\big(1-\frac{\langle
\hat{N}_{q,j^{\prime}}\rangle}{\Omega}\big) 
  \big(1-\frac{\langle \hat{N}_{q,j}\rangle}{\Omega}\big)}\sum_{\nu=1, 2}Y_{j^{\prime},\nu}Y_{j,\nu}\langle
[\bar{P}^{\dagger}_{j}, \bar{P}_{j}] \rangle, \cr
  &=&\sum_{\nu=1, 2}Y_{j^{\prime},\nu}Y_{j,\nu}\sqrt{\big(1-\frac{\langle \hat{N}_{q,j^{\prime}}\rangle}
  {\Omega}\big)\big(1-\frac{\langle \hat{N}_{q,j}\rangle}{\Omega}\big)}.
\eea\ewt
We now will explain how we express the occupation number for each level $j$ as function of the RPA
amplitudes. We start with expectation value of~(\ref{a44}) in the RPA state, we can write
\bea
\langle \hat{N}_{q,j}\rangle \simeq 2\langle P^{\dagger}_{j} P_{j} \rangle+\frac{2}{\Omega -1}\langle
{P^{\dagger}_{j}}^{2} P_{j}^{2} \rangle. 
\eea
Using~(\ref{a20}) and~(\ref{a23}) we find,
\beq
\langle {P^{\dagger}_{j}}^{2} P_{j}^{2} \rangle = K_{j, 1}+K_{j, 2}\langle P^{\dagger}_{j} P_{j} \rangle +
K_{j, 3} \langle P_{j} P^{\dagger}_{j} \rangle +K_{j, 4} \langle P^{\dagger}_{j} P^{\dagger}_{j} \rangle
\eeq
where,
\bea
K_{j, 1}&=&(1+\frac{2}{\Omega})\bigg\{\big(X_{j,1}Y_{j,1}+X_{j,2}Y_{j,2}\big)^{2}
+2\big(Y_{j,1}^{2}+Y_{j,2}^{2}\big)^{2}\bigg\},\cr
K_{j, 2}&=&-\frac{2}{\Omega^{2}}\bigg\{2\big(X_{j,1}Y_{j,1}+X_{j,2}Y_{j,2}\big)^{2}
+3\big(Y_{j,1}^{2}+Y_{j,2}^{2}\big)^{2}\bigg\},\cr
K_{j, 3}&=&-\frac{2}{\Omega^{2}}\bigg\{\big(X_{j,1}Y_{j,1}+X_{j,2}Y_{j,2}\big)^{2}
+3\big(Y_{j,1}^{2}+Y_{j,2}^{2}\big)^{2}\bigg\},\cr
K_{j, 4}&=&-\frac{6}{\Omega^{2}}\big(X_{j,1}Y_{j,1}+X_{j,2}Y_{j,2}\big)
\big(Y_{j,1}^{2}+Y_{j,2}^{2}\big)
\eea
and,
\bea
\langle P^{\dagger}_{j} P_{j} \rangle &=&(Y_{j,1}^{2}+Y_{j,2}^{2})(1-\frac{\langle
\hat{N}_{q,j} \rangle}{\Omega}), \cr
\langle P^{\dagger}_{j} P^{\dagger}_{j} \rangle&=&(X_{j,1}Y_{j,1}+X_{j,2}Y_{j,2})(1-\frac{\langle
\hat{N}_{q,j} \rangle}{\Omega}), \cr \cr
\langle P_{j} P^{\dagger}_{j} \rangle&=&(X_{j,1}^{2}+X_{j,2}^{2})(1-\frac{\langle
\hat{N}_{q,j} \rangle}{\Omega}).
\eea
Explicitly, we obtain
\bwt\bea
\langle \hat{N}_{q,j}\rangle &\simeq&2(Y_{j,1}^{2}+Y_{j,2}^{2})(1 -\frac{\langle \hat{N}_{q,j} \rangle}{\Omega})
+\frac{2}{(\Omega -1)}K_{1}+\frac{2}{(\Omega-1)}
\bigg\{K_{2}(Y_{j,1}^{2}+Y_{j,2}^{2})+K_{3}(X_{j,1}^{2}+X_{j,2}^{2}) \cr
&+&K_{4}(X_{j,1}Y_{j,1}+X_{j,2}Y_{j,2})\bigg\}(1-\frac{\langle\hat{N}_{q,j} \rangle}{\Omega}).
\eea\ewt
Therefore, we can express $\langle \hat{N}_{q,j}\rangle$ as function of the RPA amplitudes
\bwt\beq
\langle \hat{N}_{q,j}\rangle \simeq\frac{2(Y_{j,1}^{2}+Y_{j,2}^{2})+\frac{2}{(\Omega -1)}
\bigg\{K_{1}+\Omega\big(K_{2}(Y_{j,1}^{2}+Y_{j,2}^{2})+K_{3}(X_{j,1}^{2}+X_{j,2}^{2})
+K_{4}(X_{j,1}Y_{j,1}+X_{j,2}Y_{j,2})\big)\bigg\}}{1+\frac{2}{\Omega}(Y_{j,1}^{2}+Y_{j,2}^{2})
+\frac{2}{(\Omega -1)}\bigg\{K_{2}(Y_{j,1}^{2}+Y_{j,2}^{2})+K_{3}(X_{j,1}^{2}+X_{j,2}^{2})
+K_{4}(X_{j,1}Y_{j,1}+X_{j,2}Y_{j,2})\bigg\}}.
\eeq\ewt
In the same way, we express $\langle \hat{N}^{2}_{q,j}\rangle $ and 
$\langle \hat{N}_{q,1}\hat{N}_{q,-1}\rangle $ as follows,
\beq
\langle \hat{N}^{2}_{q,j}\rangle = 2(\Omega +1) \langle \hat{N}_{q,j}\rangle
-4\Omega(Y^{2}_{j,1}+Y^{2}_{j,2})(1-\frac{\langle\hat{N}_{q,j} \rangle}{\Omega})
\eeq
and,
\bea
\langle \hat{N}_{q,1}\hat{N}_{q,-1}\rangle &\simeq& 4\langle P^{\dagger}_{1}P_{1}P^{\dagger}_{-1}P_{-1}\rangle \cr
                    &\simeq& \frac{4 M}{1-\frac{4}{\Omega^{2}}(Y^{2}_{1,1}+Y^{2}_{1,2})
(Y^{2}_{-1,1}+Y^{2}_{-1,2})}
\eea
where $M$ is a constant depending of the RPA amplitudes, it is given by
\bwt\bea
M&=&(Y^{2}_{1,1}+Y^{2}_{1,2})(Y^{2}_{-1,1}+Y^{2}_{-1,2})+(1+\frac{2}{\Omega})\bigg\{(Y_{-1,1}X_{1,1}
+ Y_{-1,2}X_{1,2})(X_{-1,1}Y_{1,1}+X_{-1,2}Y_{1,2}) + (Y_{-1,1}Y_{1,1} \cr
&+&Y_{-1,2}Y_{1,2})(X_{-1,1}X_{1,1}+
X_{-1,2}X_{1,2})\bigg\}-\frac{2}{\Omega}\bigg\{(Y_{-1,1}X_{1,1}+Y_{-1,2}X_{1,2})(X_{-1,1}X_{1,1} 
+ X_{-1,2}X_{1,2})\langle P^{\dagger}_{1} P^{\dagger}_{1}\rangle \cr
&+& (Y_{-1,1}Y_{1,1} 
+ Y_{-1,2}Y_{1,2})(X_{-1,1}Y_{1,1}+X_{-1,2}Y_{1,2}) \langle P_{1} P_{1}\rangle  
+\bigg\{(Y_{-1,1}X_{1,1}+Y_{-1,2}X_{1,2})(X_{-1,1}Y_{1,1} \cr
&+&X_{-1,2}Y_{1,2})+(Y_{-1,1}Y_{1,1} 
+ Y_{-1,2}Y_{1,2})(X_{-1,1}X_{1,1}+X_{-1,2}X_{1,2})\bigg\}
(2\langle P^{\dagger}_{1} P_{1}\rangle +\langle P_{1} P^{\dagger}_{1}\rangle) \cr 
&+&(Y_{-1,1}^{2}+Y_{-1,2}^{2})(X_{-1,1}Y_{1,1}+X_{-1,2}Y_{1,2}) \langle P^{\dagger}_{1} P^{\dagger}_{-1}\rangle
+(Y_{-1,1}^{2}+Y_{-1,2}^{2})(Y_{-1,1}Y_{1,1}+Y_{-1,2}Y_{1,2})\langle P^{\dagger}_{1}P_{-1}\rangle \cr
&+& \frac{1}{2}(Y_{1,1}^{2}+Y_{1,2}^{2})(Y_{-1,1}^{2}+Y_{-1,2}^{2})
(\langle \hat{N}_{q,1}\rangle + \langle\hat{N}_{q,-1}\rangle)\bigg\}.
\eea\ewt

\section{Method for Calculation of $\hat{N}^{k}_{q,j}$}
\label{Number}
In this Appendix, we will present our method inspired from~\cite{b21} (see also~\cite{b33}) for the derivation 
of the quantities of type $\hat{N}^{k}_{q,j}$ in the case of the two-levels pairing model. At first, we recall 
that in this model, the operators $\hat{N}_{q,j}$, $P^{\dagger}_{q,j}$ and $P_{q,j}$ close the $SU(2)$ algebra 
for each level $j$. Consequently the two level model fulfills an $SU(2)\times SU(2)$ algebra. Thanks to this 
special groupe structure, it is easy to find an complete ortho-normalized basis for the Hilbert subspace
corresponding to each level $j$,
\beq
|n_{j}\rangle =\sqrt{\frac{\Omega^{n_{j}}(\Omega-n_{j})!}
{\Omega!n_{j}!}}{P^{\dagger}_{j}}^{n_{j}}|-\rangle,\; j=\pm 1
\label{h1}
\eeq
where $n_{j}=0,1,\ldots,\Omega$.
Using this basis, we can express the identity operator relatively to each level $j$ as 
\bwt\beq
1_{j}=\sum_{n_{j}=0}^{\Omega}|n_{j}\rangle \langle n_{j}|=|-\rangle \langle
-|+\sum_{n_{j}=1}^{\Omega}\frac{\Omega^{n_{j}}(\Omega-n_{j})!}{\Omega!n_{j}!}{P^{\dagger}_{j}}^{n_{j}}|-\rangle
\langle -|
P^{n_{j}}_{j}, \; j=\pm 1
\label{h2}
\eeq\ewt
therefore, we can express the projector $|-\rangle \langle -|$ as follows
\beq
|-\rangle \langle -| = 1_{j} -
\sum_{n_{j}=1}^{\Omega}\frac{\Omega^{n_{j}}(\Omega-n_{j})!}{\Omega!n_{j}!}{P^{\dagger}_{j}}^{n_{j}}
|-\rangle \langle -| P^{n_{j}}_{j}.
\label{h3}
\eeq
One sees that~(\ref{h3}) produces an expansion of the form 
\beq
|-\rangle\langle -|=\sum_{m_{j}=0}^{\Omega}\beta_{m_{j}}{P^{\dagger}_{j}}^{m_{j}}P^{m_{j}}_{j},
\label{h4}
\eeq
if we substitute~(\ref{h4}) in both lhs and rhs of~(\ref{h3}), we obtain the coefficients $\beta_{m_{j}}$
\beq
\beta_{0}=1, \;\;
\beta_{m_{j}}=-\sum_{l=0}^{m_{j}-1}\frac{\Omega^{m_{j}-l}(\Omega-m_{j}+l)!}{\Omega!(m_{j}-l)!}\beta_{l}.
\label{h5}
\eeq
For example, the first terms $\beta_{m_{j}}$ are explicitly given by,
\bea
\beta_{0}&=& 1,\cr
\beta_{1}&=&-1,\cr
\beta_{2}&=& \frac{\Omega -2}{2(\Omega -1)},\cr
\beta_{3}&=&-\frac{\Omega^{2}-6\Omega +12}{6(\Omega -1)(\Omega -2)}.
\label{h6}
\eea
However, to calculate the quantities $\hat{N}_{q,j}$ and $\hat{N}^{2}_{q,j}$, one can expand these operators as
\beq
\hat{N}^{k}_{q,j}=\sum_{l_{j}=1}^{\Omega}\alpha^{(k)}_{l_{j}}{P^{\dagger}_{j}}^{l_{j}}P^{l_{j}}_{j},
\; j=\pm 1.
\label{h7}
\eeq
For all operators of the form $ \hat{N}^{k}_{q,j}$, using the fact that
\beq
\hat{N}_{q,j}|n_{j}\rangle = 2 n_{j}|n_{j}\rangle,
\label{h8}
\eeq
we can calculate
\bwt\beas
\hat{N}^{k}_{q,j} &=& \sum_{n_{j}=0}^{\Omega}\hat{N}^{k}_{q,j}|n_{j}\rangle \langle n_{j}| \slabel{hh1}\\
 &=& \sum_{n_{j}=0}^{\Omega}(2n_{j})^{k}|n_{j}\rangle \langle n_{j}| \slabel{hh2}\\
 &=& \sum_{n_{j}=0}^{\Omega}\frac{\Omega^{n_{j}}(\Omega-n_{j})!}{\Omega!n_{j}!}
 (2n_{j})^{k}{P^{\dagger}_{j}}^{n_{j}}|-\rangle \langle -| P^{n_{j}}_{j} \slabel{hh3}.
\label{h9}
\eeas\ewt
By inserting of~(\ref{h4}) in the rhs~(\ref{hh3}) and substituting~(\ref{h7}) into lhs of~(\ref{hh3}), we obtain
\bwt\bea
\hat{N}^{k}_{q,j} &=&
\sum_{n_{j}=0}^{\Omega}\frac{\Omega^{n_{j}}(\Omega-n_{j})!}{\Omega!n_{j}!}(2n_{j})^{k}\sum_{m_{j}=0}^{\Omega}
\beta_{m_{j}}{P^{\dagger}_{j}}^{n_{j}} {P^{\dagger}_{j}}^{m_{j}}P^{m_{j}}_{j} P^{n_{j}}_{j} \cr
\sum_{l_{j}=1}^{\Omega}\alpha^{(k)}_{l_{j}}{P^{\dagger}_{j}}^{l_{j}}P^{l_{j}}_{j}&=&
\sum_{n_{j}=1}^{\Omega}\frac{\Omega^{n_{j}}(\Omega-n_{j})!}{\Omega!n_{j}!}(2n_{j})^{k}\sum_{m_{j}=0}^{\Omega}
\beta_{m_{j}}{P^{\dagger}_{j}}^{n_{j}+m_{j}}P^{n_{j}+m_{j}}_{j} \cr
&=&\sum_{l_{j}=1}^{\Omega}\sum_{m_{j}=0}^{l_{j}-1}\frac{\Omega^{l_{j}-m_{j}}(\Omega-l_{j}+m_{j})!}
{\Omega!(l_{j}-m_{j})!}(2(l_{j}-m_{j}))^{k}\beta_{m_{j}}{P^{\dagger}_{j}}^{l_{j}}P^{l_{j}}_{j}.
\label{h10}
\eea\ewt
Therefore, by identification, from~(\ref{h10}) it is easy to get the coefficients $\alpha^{(k)}_{l_{j}}$,
\bwt\beq
\alpha^{(k)}_{l_{j}}=\sum_{m_{j}=0}^{l_{j}-1}\frac{\Omega^{l_{j}-m_{j}}(\Omega-l_{j}+m_{j})!}
{\Omega!(l_{j}-m_{j})!}(2(l_{j}-m_{j}))^{k}\beta_{m_{j}}.
\label{h11}
\eeq\ewt
To calculate $\hat{N}_{q,j}$ we put $k=1$ in~(\ref{h7}), and find  
\beq 
\hat{N}_{q,j}=\sum_{l_{j}=1}^{\Omega}\alpha^{(1)}_{l_{j}}{P^{\dagger}_{j}}^{l_{j}}P^{l_{j}}_{j}, 
\label{h12}
\eeq
with,
\beq
\alpha^{(1)}_{l_{j}}=\sum_{m_{j}=0}^{l_{j}-1}\frac{\Omega^{l_{j}-m_{j}}(\Omega-l_{j}+m_{j})!}
{\Omega!(l_{j}-m_{j})!}(2(l_{j}-m_{j}))\beta_{m_{j}}.
\label{h13}
\eeq
The first coefficients $\alpha^{(1)}_{l_{j}}$ are explicitly given by,
\bea
\alpha^{(1)}_{1}&=&2, \cr
\alpha^{(1)}_{2}&=& \frac{2}{\Omega -1}, \cr
\alpha^{(1)}_{3}&=& \frac{4}{(\Omega -1)(\Omega -2)}, \cr
\alpha^{(1)}_{4}&=& \frac{2(5\Omega -6)}{(\Omega -1)^{2}(\Omega -2)(\Omega -3)}.
\label{h14}
\eea
Therefore, for $ \hat{N}_{q,j}$ we can write 
\bwt\beq 
\hat{N}_{q,j}=2 P^{\dagger}_{j}P_{j} + \frac{2}{\Omega -1}{P^{\dagger}_{j}}^{2}P_{j}^{2} + 
\frac{4}{(\Omega -1)(\Omega -2)}{P^{\dagger}_{j}}^{3}P_{j}^{3} +
\frac{2(5\Omega -6)}{(\Omega -1)^{2}(\Omega -2)(\Omega -3)}{P^{\dagger}_{j}}^{4}P_{j}^{4} + \ldots.
\label{h15}
\eeq\ewt
In the same way, it is very easy with this method to find an expansion for $ \hat{N}^{2}_{q,j}$, it is sufficient 
to put $k=2$ in~(\ref{h7}) and calculate $\alpha^{(2)}_{l_{j}}$. We find
\bwt\beq 
\hat{N}^{2}_{q,j}=4 P^{\dagger}_{j}P_{j} + \frac{4(\Omega +1)}{\Omega -1}{P^{\dagger}_{j}}^{2}P_{j}^{2} + 
\frac{8(\Omega +1)}{(\Omega -1)(\Omega -2)}{P^{\dagger}_{j}}^{3}P_{j}^{3} +
\frac{4(\Omega +1)(5\Omega -6)}{(\Omega -1)^{2}(\Omega -2)(\Omega -3)}{P^{\dagger}_{j}}^{4}P_{j}^{4} +\ldots.
\label{h16}
\eeq\ewt   


\newpage

\begin{figure}                       
\resizebox{0.65 \columnwidth}{!}{\includegraphics{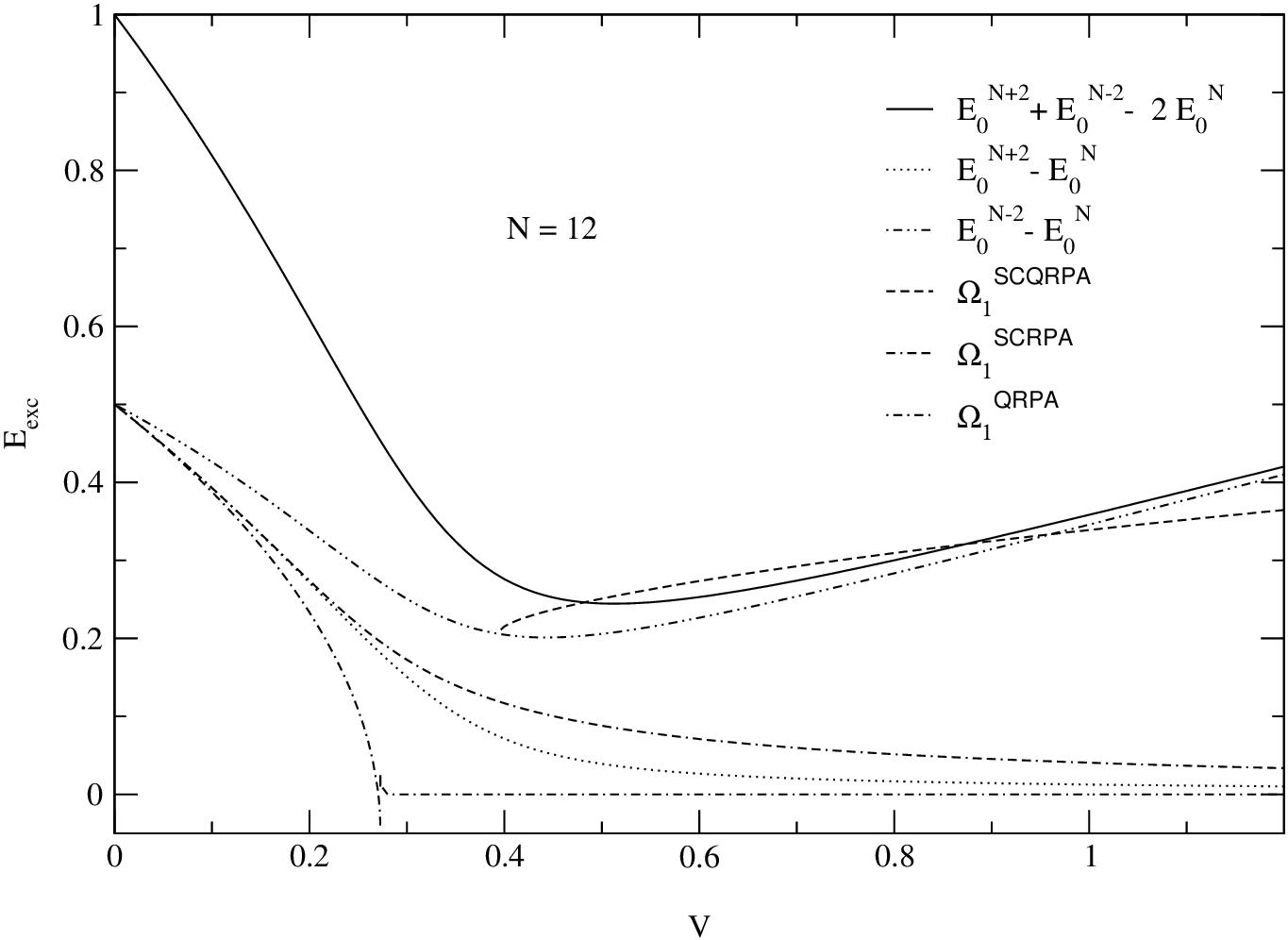}}
\caption{Excitation energy of the soft (spurious) mode (energies are divided by $2 \epsilon$) 
as a function of the variable $V=g\Omega/2\epsilon$ described in the text and for particle number $N=12$. 
The spin of the levels is $J=11/2$. The results refer to exact calculations (solid line, dotted line 
and double-dot dashed line), RPA and QRPA (dot-dashed line), SCQRPA (dashed line) and SCRPA (double-dash 
dotted line).}
\label{eigen12}
\end{figure} 

\end{document}